\documentclass[usenatbib,usegraphics]{mn2e}                 %
\def\kms{km\,s$^{-1}$}

\def\M{M$_{\odot}$}

\def\ebv{$E(B-V)$}
\def\rv{$R_{V}$}
\def\av{$A_{V}$}
\def\dm15{$\Delta m_{15}(B)$}
\def\psec{\ensuremath{\,.\!\!^{s}}}
\def\parcmin{\ensuremath{\,.\!\!\arcmin}}
\def\parcsec{\ensuremath{\,.\!\!\arcsec}}
\def\dijd{$\Delta t_{max}(I)$}

\def\CI{C\,{\sc i}}

\def\SiII{Si\,{\sc ii}}

\def\MgII{Mg\,{\sc ii}}
\def\CaII{Ca\,{\sc ii}}
\def\FeII{Fe\,{\sc ii}}

\def\CoII{Co\,{\sc ii}}

\def\HeI{He\,{\sc i}}

\def\ergs{\,(erg\,s$^{-1}$)}

\input{psfig.tex}

\title[Supernova 2002cv]
{SN~2002cv: A Heavily Obscured Type Ia Supernova}

\author[Elias-Rosa et al.]
{N. Elias-Rosa$^{1,2}$, S. Benetti$^{2}$, M. Turatto$^{2}$, E.
Cappellaro$^{2}$, S. Valenti$^{3}$, A.A. Arkharov$^{4}$,
\newauthor
J. E. Beckman$^{5}$, A. Di Paola$^{6}$, M. Dolci$^{7}$, A.V.
Filippenko$^{8}$, R.J. Foley$^{8}$, K. Krisciunas$^{9}$,
\newauthor
V.M. Larionov$^{4,10}$, W. Li$^{8}$, W.P.S. Meikle$^{11}$, A.
Pastorello$^{12}$, G. Valentini$^{7}$ and
\newauthor
W. Hillebrandt$^{1}$
\\
$^1$Max-Planck-Institut f\"{u}r Astrophysik, Karl-Schwarzschild-Str. 1, D-85748 Garching bei M\"{u}nchen, Germany.\\
$^2$INAF - Osservatorio Astronomico di Padova, vicolo dell'Osservatorio 5, I-35122 Padova, Italy. \\
$^3$European Southern Observatory, Karl-Schwarzschild-Str. 2, D-85748 Garching bei M\"{u}nchen, Germany.\\
$^4$Main (Pulkovo) Astronomical Observatory of RAS, St.
Petersburg, Russia \\
$^5$Instituto de Astrof\'isica de Canarias, C/ V\'ia L\'actea s/n,
E-38200, La Laguna, Tenerife, Spain.\\
$^6$INAF - Osservatorio Astronomico di Roma, via Frascati 33, Monteporzio Catone, I-00040 Roma, Italy.\\
$^7$INAF - Osservatorio Astronomico di Collurania, via M. Maggini, I-64100 Teramo, Italy.\\
$^8$Department of Astronomy, University of California, Berkeley, CA 94720-3411, USA.\\
$^9$Department of Physics, Texas A\&M University, College Station, TX 77843-4242, USA.\\
$^{10}$Astronomical Institute of St. Petersburg State
University, St. Petersburg, Russia\\
$^{11}$Astrophysics Group, Imperial College London, Blackett Laboratory, Prince Consort Road, London, SW7 2AZ, UK.\\
$^{12}$Astrophysics Research Centre, Queen's University Belfast,
BT7 1NN,
UK.\\
}

\date{Received ................; accepted ................}

\begin{document}

\maketitle

\begin{abstract}
We present $VRIJHK$ photometry, and optical and near-infrared
spectroscopy, of the heavily extinguished Type~Ia supernova (SN)
2002cv, located in NGC 3190, which is also the parent galaxy of
the Type Ia SN~2002bo. SN~2002cv, not visible in the blue, has a
total visual extinction of $8.74 \pm 0.21$ mag. In spite of this
we were able to obtain the light curves between $-10$ and +207
days from the maximum in the $I$ band, and also to follow the
spectral evolution, deriving its key parameters. We found the peak
$I$-band brightness to be $I_{max} = 16.57 \pm 0.10$ mag, the
maximum absolute $I$ magnitude to be $M^{max}_{I} = -18.79 \pm
0.20$, and the parameter $\Delta m_{15}(B)$ specifying the width
of the $B$-band light curve to be $1.46 \pm 0.17$ mag. The latter
was derived using the relations between this parameter and $\Delta
m_{40}(I)$ and the time interval $\Delta t_{max}(I)$ between the
two maxima in the $I$-band light curve. As has been found for
previously observed, highly extinguished SNe~Ia, a small value of
$1.59 \pm 0.07$ was obtained here for the ratio \rv\ of the
total-to-selective extinction ratio for SN~2002cv, which implies a
small mean size for the grains along the line of sight toward us.
Since it was found for SN~2002bo a canonical value of 3.1, here we
present a clear evidence of different dust properties inside NGC
3190.
\end{abstract}


\section{Introduction} \label{int}

Type Ia supernovae (SNe~Ia) are believed to result from the
explosion of white dwarfs when they reach the Chandrasekhar limit
after accreting mass from a nearby companion star. The overall
homogeneity and the small brightness dispersion at maximum qualify
SNe~Ia as the most powerful tools for measuring cosmological
distances (see \citealt{filippenko05} for a comprehensive review).
Over the last decade, a number of studies have been dedicated to
the detailed analysis of light curves of SNe~Ia and the
determination of the peak magnitudes (e.g.,
\citealt{hamuy96a,riess96,perlmutter97,phillips99,prieto06}),
host-galaxy extinctions (e.g.,
\citealt{krisc04b,wangl05,krisc06a,elias06}), statistical analysis
of spectral properties (e.g.,
\citealt{benetti05,mazzali05,branch06,hachinger06,mazzali07}), and
progenitor and explosion models
(\citealt{hillebrandt00,gamezo05,ropke06a,ropke06b,nomoto06}). All
these works have been possible only thanks to intensive campaigns
of observation and accurate calibration of nearby SNe~Ia (e.g.,
\citealt{suntzeff96,krisc03,jha06a,benetti04,pastorello06b}).

A crucial factor to be considered in the calibration of SNe is the
extinction occurring inside the host galaxy, which is generally
derived by measuring the reddening and adopting a standard value
for the total-to-selective absorption ratio, \rv\ =\av\/\ebv. On
the other hand, it is well known that \rv\ depends on the nature
of interstellar dust and varies even with the line of sight inside
the Galaxy from \rv\ $\approx$ 2 to $\sim$ 5.5
\citep{fitzpatrick04,geminale05} with an average value \rv\ = 3.1
\citep{seaton79,savage79}. Whereas very little is known for other
galaxies, usually it is assumed that the dust has the same average
properties as in the Galaxy. In most cases, measurements along the
line of sight to the SNe show canonical values of \rv\ = 3.1 (e.g.
SN~1994D - \citealt{patat96}, SN~2002bo - \citealt{benetti04},
SN~2004eo - \citealt{pastorello06b}), but for a few SNe these
measurements show evidence for extremely low values of \rv. In
particular, the Type Ia SN~1999cl \citep{krisc06a} was reddened by
dust with \rv\ = 1.55 $\pm$ 0.08. For the core-collapse SN~2002hh
\citep{pozzo06} a two-component extinction model has been proposed
to match coeval spectral templates with \av\ = 3.3 mag, \rv\ =
1.1, and with \av\ = 1.7 mag, \rv\ = 3.1, respectively. For
another heavily extinguished SN~Ia, SN~2003cg, \cite{elias06}
found \rv\ = 1.80 $\pm$ 0.19 and \ebv\ = 1.33 $\pm$ 0.11 mag.
Given the dependence of \rv\ on the dust properties, the dust
along the line of sight to these SNe seems to have very small
grain size. An alternative explanation calls for the contamination
by unresolved light echo from circumstellar dust \citep{wangl05}.
The growing number of SNe Ia associated with dust clouds with
unusual properties is of interest not only in the context of SN
calibration, but also for the debate on progenitor scenarios
\citep{patat07}.

\begin{figure*}
\centering
\psfig{figure=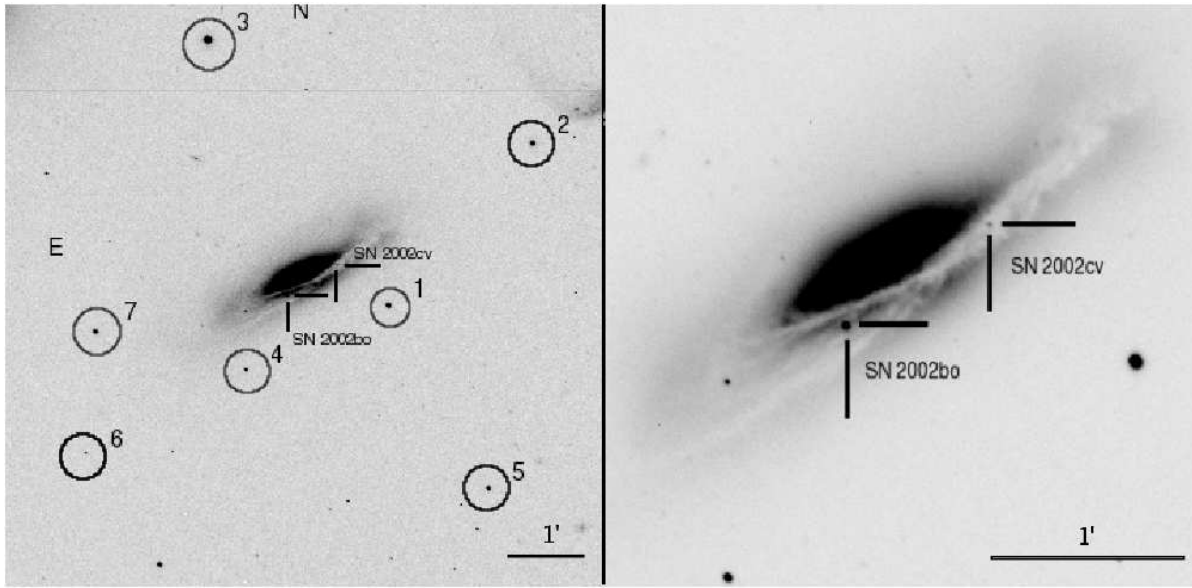,width=16cm,height=7.8cm}
\caption{$R$-band image of SN~2002cv in NGC 3190 taken with the 2.5~m
Isaac Newton Telescope + WFC on 2002 June 27 (field of view $\sim
8'\times8'$). SN~2002bo and the local sequence stars are also
indicated (see Table \ref{tabla_seq}).} \label{fig_seq}
\end{figure*}

In this paper we discuss the case of SN 2002cv, one of the most
obscured SNe ever observed \citep{dipaola02}. It was discovered in
the spiral galaxy NGC 3190 (heliocentric velocity 1271 \kms) on
2002 May 13.7 (UT dates are used throughout this paper) by
\cite{lavionov02} during the campaign of monitoring of SN~2002bo,
a SN~Ia extensively studied by \cite{benetti04} and
\cite{krisc04b}. The new source was found 18\arcsec\ W and
10\arcsec\ N of the galactic nucleus ($\alpha =
10^{h}18^{m}03\psec68, \delta = +21^{o}50\arcmin06\parcsec20$,
J2000.0) projected on the galaxy dust lane (Figure \ref{fig_seq}).
An early optical spectrum \citep{turatto02} showed a very red
continuum with almost no signal blueward of 600 nm, which was
attributed to very high extinction. Infrared spectra taken at the
United Kingdom Infrared Telescope (UKIRT) on May 22.3 and 23.3
\citep{meikle02} suggested that SN~2002cv was a SN~Ia similar to
the bright SN~1991T \citep{filippenko97,filippenko92a}. This
classification was confirmed by \cite{filippenko02}.
\cite{meikle02} also estimated that the value of \av\ toward
SN~2002cv actually exceeded 6 mag, the highest extinction ever
recorded for a SN~Ia.

Here we present a wide set of photometric data obtained
worldwide for SN~2002cv along with selected spectroscopic
observations. The description of the data reduction and the
analysis of the light curves are given in Section \ref{ph}.
Section \ref{spec} presents the optical and IR spectra. In Section
\ref{redd} we discuss the reddening, while in Section \ref{param}
we derive the main parameters of the SN using empirical relations
both from the literature and derived here. We conclude
with a summary of the results (Section \ref{summary}).

\section{Photometry} \label{ph}
\subsection{Data acquisition and reduction} \label{ph_obs}

The fact that another SN~Ia, SN~2002bo, exploded in the same
galaxy a few months before SN~2002cv, makes available pre-discovery
optical and near-infrared (NIR) data and hence templates for
photometric measurements via image subtraction. SN~2002cv was
observed for seven months, from day $-10.7$ to +205.9 relative to
$I$-band maximum light. Here we present the data collected by four
different teams: University of California, Berkeley (USA),
Imperial College (UK), Padova Observatory (INAF, Italy), and Teramo
Observatory (INAF, Italy). A total of 9 different instrument
configurations was used:

\begin{enumerate}

\item 1~m Nickel telescope (Lick Observatory, Mt. Hamilton,
California, USA), equipped with a SITe thinned CCD ($1024 \times
1024$ pixels, 0.28\arcsec pix$^{-1}$, field of view $6\parcmin3
\times 6\parcmin3$);

\item 0.72~m Teramo-Normale Telescope (Terano, Italy), equipped
with a TK512CB1-1 CCD ($512 \times 512$ pixels, 0.46\arcsec
pix$^{-1}$, field of view $3\parcmin92 \times 3\parcmin92$);

\item 0.76~m Katzman Automatic Imaging Telescope (Lick Observatory,
Mt. Hamilton, California, USA), equipped with a SITe AP7 CCD
($512 \times 512$ pixels, 0.8\arcsec pix$^{-1}$, field of view
$6\parcmin7 \times  6\parcmin7$);

\item 1.82~m Copernico telescope of Mt. Ekar (Asiago, Italy),
equipped with AFOSC (thinned TEK CCD, $1024 \times 1024$ pixels,
0.473\arcsec pix$^{-1}$, field of view $8\parcmin14 \times
8\parcmin14$);

\item 3.6~m ESO/NTT telescope (La Silla, Chile), equipped with EMMI
(Tektronix TK1034 CCD, $1024 \times 1024$ pixels; blue arm 0.37\arcsec
pix$^{-1}$, field of view $6\parcmin2 \times 6\parcmin2$; red arm
0.167\arcsec pix$^{-1}$, field of view $9\parcmin1 \times 9\parcmin9$);

\item 2.5~m Isaac Newton Telescope (Roque de los Muchachos
Observatory, La Palma, Spain), equipped with Wide Field Camera (4
thinned EEV CCDs, 2048 $\times$ 4096 pixels, 0.33\arcsec pix$^{-1}$,
field of view $34\arcmin \times 34\arcmin$);

\item 1.0~m Jacob Kapteyn Telescope (Roque de los Muchachos
Observatory, La Palma, Spain), equipped with JAG (SITe2 CCD,
$2048 \times 2048$ pixels, 0.33\arcsec pix$^{-1}$, field of view
$10\arcmin \times 10\arcmin$);

\item 3.6~m ESO/NTT Telescope (La Silla, Chile), equipped with SofI
(NIR HgCdTe Hawaii array, $1024 \times 1024$ pixels, 0.288\arcsec
pix$^{-1}$, field of view $4\parcmin94 \times 4\parcmin94$);

\item 1.08~m AZT-24 telescope (Campo Imperatore, Italy), equipped
with SWIRCAM (NIR HgCdTe PICNIC array, $256 \times 256$ pixels,
1.04\arcsec pix$^{-1}$, field of view $4\parcmin4 \times 4\parcmin4$).

\end{enumerate}

The photometric observations were processed using
IRAF\footnote{IRAF is written and supported by the IRAF
programming group at the National Optical Astronomy Observatories
(NOAO) in Tucson, Arizona, which are operated by the Association
of Universities for Research in Astronomy (AURA), Inc. under
cooperative agreement with the National Science Foundation.}
routines with the standard recipe for CCD images (trimming,
overscan, bias, and flat-field corrections). For the NIR bands, we
also performed sky image subtraction, as well as image coaddition
to improve the signal-to-noise ratio (S/N).

In all cases, because of the strong luminosity gradient at the SN
position, we made use of the template subtraction technique in order
to measure the SN magnitudes \citep{filippenko86}.
Image subtraction was performed
using the ISIS program \citep{alard00}. As reference images
(templates) of the host galaxy, we selected those taken about two
months before SN~2002cv discovery with the Asiago 1.82~m Copernico
telescope + AFOSC on 2002 March 21, and the AZT-24 Telescope +
SWIRCAM on 2002 March 28, for the optical and NIR bands,
respectively. After geometrical and photometric registration of
the two images (target image and template) and the degradation of
the image with the best seeing to match the worst one, the
template was subtracted from the target image. Hereafter, the
instrumental magnitude of the SN was measured with the
point-spread function (PSF) fitting technique using the DAOPHOT
package on the subtracted image. Reference stars in the SN field
(see Figure \ref{fig_seq}) were also measured using the IRAF PSF
fitting routine on the original image.

In order to calibrate the instrumental magnitudes onto a standard
photometric system, we used the specific colour-term equations for
each of the various instrumental configurations. For the optical
bands, these were derived from observations during photometric
nights of several standard fields \citep{land92}. In turn, the
photometric zero-points for non-photometric nights were determined
using the magnitudes of the local sequence stars in the SN field
(Table \ref{tabla_seq}). For the NIR, because of the small number of
standard stars observed each night, we used the average colour
terms provided by the telescope teams.

Actually, after the detailed monitoring of SN~2002bo, accurate
estimates are available for the magnitudes of a number of local
standard stars which are used to calibrate our data
(\citealt{benetti04}; \citealt{krisc04b}; KAIT observations).
The order of the local sequence is that given by Benetti et al.
(2004; see their Figure 1 and Table 1) with the addition of star
number 7 which corresponds to star number 3 of Krisciunas et al.
(2004b; see their Figure 1{\it e} and Table 1). For the NIR
magnitudes we calibrated two stars of the local sequence during
photometric nights. The NIR magnitudes of the single star in
common with \cite{krisc04b} agree to better than 0.02 mag.

We also included the data from \cite{dipaola02} (see Section
\ref{ph_lc}) obtained using the AZT-24 Telescope of the Campo
Imperatore Observatory, equipped with the NIR camera
SWIRCAM. These data were re-calibrated against our local sequence,
while for a few deviant points a new measurement was necessary.

Whereas standard colour corrections properly bring the magnitudes
of normal stars to the standard system, it is known that this
does not work well for SNe because of their specific spectral
energy distribution (SED). This and the differences between
bandpasses cause a significant dispersion of the photometric
measurements obtained with different instruments. We note that
during the follow-up observations of SN~2002cv, we used seven
different instruments for the optical and two for the NIR
observations.

The accurate SN calibration procedure, usually called S-correction
(\citealt{suntzeff00}; \citealt{stritz02}; \citealt{krisc03};
\citealt{pignata04a}; \citealt{pignata04b}), requires that the SED
of the object and the response curve of the instruments used for
the observations are both accurately known.

We computed the S-corrections for the $VRI$ bands\footnote{We did
not compute the S-correction for NIR bands and for the late time
optical observations because of the lack of suitable spectra.} by
using the flux-calibrated spectra of SN~2002cv (Section
\ref{spec}). Since our spectra did not cover all photometric
epochs, we completed the spectral database by adding spectra of
unreddened normal SNe~Ia such as SN~1992A (\citealt{suntzeff96}),
SN~1994D (\citealt{patat96}), and SN~1996X (\citealt{salvo01}),
properly reddened to match those of SN~2002cv (see Section
\ref{redd} for more details).

\begin{figure}
\psfig{figure=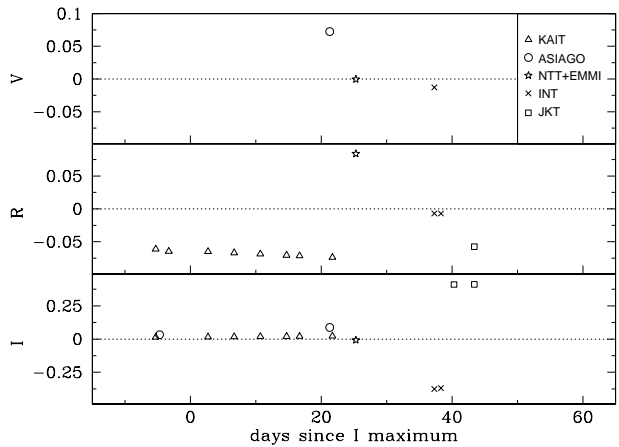,angle=270,width=9.5cm,height=7.1cm}
\caption{Summary of the S-corrections derived for the $VRI$ bands of
the different instruments (see legend) at early times. These
corrections have been added to the first-order corrected SN~2002cv
magnitudes to convert them to the standard system. The dotted line
shows the zero correction.} \label{fig_scorr1}
\end{figure}

\begin{figure}
\psfig{figure=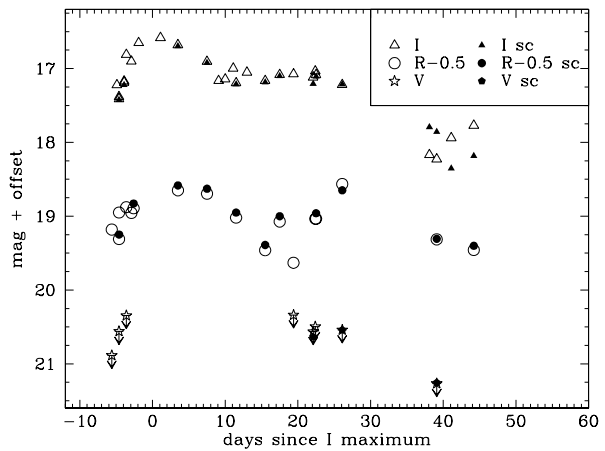,angle=270,width=9.5cm,height=7.1cm}
\caption{Comparison between the original $VRI$ light curves
(empty symbols) of SN~2002cv and the corrected ones (filled
symbols). Note that we were not able to apply the S-correction to
some instruments; ``sc" stands for ``S-corrected."} \label{fig_scorr2}
\end{figure}

The corrections for SN~2002cv are in general relatively small
($\leq0.1$ mag), as seen in Figure \ref{fig_scorr1} and Table
\ref{tabla_scorrection_ph_opt}, except for the $I$ band of the INT
and JKT, where the corrections are as high as 0.40 mag. These
telescopes use Sloan Gunn $i$ and Harris $I$ filters, respectively,
which differ significantly from the Bessell ones. The SN
magnitudes, calibrated using this technique, agree fairly well
(Figure \ref{fig_scorr2}). Note that the data from TNT and the Lick
Nickel 1~m telescope were not corrected because some of the
required instrumental information was not available. In any case,
the measurements from these instruments appear to be in good
agreement with the S-corrected photometry from other
instruments (see Figure \ref{fig_scorr2}).\\

Uncorrected and S-corrected optical magnitudes are reported in
Tables \ref{tabla_noscor_ph_opt} and \ref{tabla_ori_ph_opt},
respectively, and NIR measurements are listed in Table
\ref{tabla_ori_ph_ir}. Magnitudes are presented together with
their uncertainties, which were computed as the quadrature sum
of the following contributions: the square root of the PSF fitting
errors on the subtracted images, calibration errors (root-mean square
[rms] of the observed magnitudes of the local sequence stars), and
errors associated to the S-correction (rms deviation with respect to
the low-order polynomial fit over phase).

\begin{table*}
\centering
\caption{Adopted magnitudes for the local sequence stars,
coded as in Figure
\ref{fig_seq} (see Section \ref{ph_obs}).} \label{tabla_seq}
\begin{tabular}{ccccccc}
\hline
~~star~~ & $V$ & $R$ & $I$ & $J$ & $H$ & $K$ \\
\hline
1        &$14.42\pm0.02$ &$13.99\pm0.03$ &$13.58\pm0.01$ &$13.05\pm0.01$ &$12.71\pm0.01$ &$12.67\pm0.01$\\
2        &$14.33\pm0.04$ &$13.93\pm0.04$ &$13.57\pm0.02$ &       -       &       -       &       -      \\
3        &$12.40\pm0.05$ &$12.05\pm0.05$ &$11.68\pm0.05$ &       -       &       -       &       -      \\
4        &$17.20\pm0.02$ &$16.35\pm0.03$ &$15.65\pm0.03$ &$14.83\pm0.01$ &$14.20\pm0.01$ &$14.12\pm0.01$\\
5        &$15.64\pm0.04$ &$15.26\pm0.04$ &$14.91\pm0.01$ &       -       &       -       &       -      \\
6        &$17.90\pm0.02$ &$17.51\pm0.05$ &$17.05\pm0.10$ &       -       &       -       &       -      \\
7        &$14.92\pm0.01$ &$14.49\pm0.01$ &$14.06\pm0.01$ &       -       &       -       &       -      \\
\hline
\end{tabular}
\end{table*}

\begin{table*}
\centering
\caption{S-correction to be added to the data of SN~2002cv in
Table \ref{tabla_noscor_ph_opt}.} \label{tabla_scorrection_ph_opt}
\begin{tabular}{lcrcccccc}
\hline
   date  &   JD $-$   & Phase*&  $V$   &  $R$   &  $I$    & Instr.\\
         & 2,400,000.00 &(days)&              &                  &                  &       \\
\hline
15/05/02 & 52409.75 &  $-$5.3  &        -     & $-$0.061(0.009) &  0.014(0.011) & KAIT \\
15/05/02 & 52410.40 &  $-$4.7  &        -     &         -     &  0.033(0.019) & EKAR \\
17/05/02 & 52411.75 &  $-$3.3  &        -     & $-$0.064(0.009) &          -    & KAIT \\
23/05/02 & 52417.75 &   2.7  &        -     & $-$0.065(0.009) &  0.016(0.011) & KAIT \\
27/05/02 & 52421.75 &   6.7  &        -     & $-$0.067(0.009) &  0.017(0.011) & KAIT \\
31/05/02 & 52425.75 &  10.7  &        -     & $-$0.069(0.009) &  0.018(0.011) & KAIT \\
04/06/02 & 52429.75 &  14.7  &        -     & $-$0.071(0.009) &  0.020(0.011) & KAIT \\
06/06/02 & 52431.75 &  16.7  &        -     & $-$0.071(0.009) &  0.020(0.011) & KAIT \\
10/06/02 & 52436.37 &  21.3  & 0.073(0.044) &         -     &  0.088(0.019) & EKAR \\
11/06/02 & 52436.75 &  21.7  &        -     & $-$0.074(0.009) &  0.022(0.011) & KAIT \\
14/06/02 & 52440.39 &  25.3  & $-$0.001(0.017) &  0.084(0.019) & -0.007(0.012) & EMMI \\
26/06/02 & 52452.40 &  37.3  & $-$0.013(0.003) & $-$0.007(0.001) & -0.376(0.053) & INT  \\
27/06/02 & 52453.42 &  38.3  &        -     & $-$0.007(0.001) & -0.369(0.053) & INT \\
29/06/02 & 52455.43 &  40.3  &        -     &         -     &  0.411(0.028) & JKT \\
02/07/02 & 52458.45 &  43.4  &        -     & $-$0.057(0.006) &  0.412(0.028) & JKT \\
\hline
\end{tabular}
\begin{flushleft}
*Relative to $I_{max}$ (JD = 2,452,415.09).\\
KAIT = 0.76~m Katzman Automatic Imaging Telescope + CCD,
0.80\arcsec pix$^{-1}$; EKAR = 1.82~m Copernico telescope + AFOSC,
0.47\arcsec pix$^{-1}$; EMMI = 3.6~m ESO NTT + EMMI, 0.167\arcsec pix$^{-1}$;
INT = 2.5~m Isaac Newton Telescope + WFC, 0.33\arcsec pix$^{-1}$; JKT = 1.0~m
Jacob Kapteyn Telescope + JAG, 0.33\arcsec pix$^{-1}$.\\
\end{flushleft}
\end{table*}

\begin{table*}
\centering
\caption{Original optical photometry of SN 2002cv.}
\label{tabla_noscor_ph_opt}
\begin{tabular}{lcrcccccc}
\hline
   UT Date  &   JD $-$   & Phase*&  $V$  &   $R$   &  $I$        & Instr.\\
         & 2,400,000.00 &(days)&              &                  &                  &       \\
\hline
14/05/02 & 52408.75 & $-$6.3  & $\geq20.97$ & 19.81(0.18) &       -     & N1mT\\
14/05/02 & 52409.37 & $-$5.7  & $\geq20.48$ &      -      & 17.23(0.14) & TNT\\
15/05/02 & 52409.74 & $-$5.4  & $\geq20.69$ & 19.62(0.27) & 17.45(0.11) & N1mT\\
15/05/02 & 52409.75 & $-$5.3  &       -     & 19.79(0.26) & 17.48(0.03) & KAIT\\
15/05/02 & 52410.39 & $-$4.7  &       -     &       -     & 17.18(0.44) & TNT\\
15/05/02 & 52410.40 & $-$4.7  &       -     &       -     & 17.28(0.04) & EKAR\\
16/05/02 & 52410.74 & $-$4.4  & $\geq20.50$ & 19.54(0.06) & 16.89(0.11) & N1mT\\
16/05/02 & 52411.41 & $-$3.7  &       -     & 19.42(0.68) & 16.69(0.18) & TNT\\
17/05/02 & 52411.75 & $-$3.3  &       -     & 19.35(0.09) &       -     & KAIT\\
17/05/02 & 52412.38 & $-$2.7  &       -     &       -     & 16.65(0.44) & TNT\\
20/05/02 & 52415.39 &  0.3  &       -     &       -     & 16.67(0.09) & TNT\\
23/05/02 & 52417.75 &  2.7  &       -     & 19.30(0.25) & 16.84(0.12) & KAIT\\
27/05/02 & 52421.75 &  6.7  &       -     & 19.34(0.18) & 17.04(0.14) & KAIT\\
28/05/02 & 52423.37 &  8.3  &       -     &       -     & 17.25(0.12) & TNT \\
29/05/02 & 52424.33 &  9.2  &       -     &       -     & 17.23(0.10) & TNT\\
30/05/02 & 52425.36 & 10.3  &       -     &       -     & 17.24(0.10) & TNT\\
31/05/02 & 52425.75 & 10.7  &       -     & 19.65(0.07) & 17.33(0.30) & KAIT\\
01/06/02 & 52427.28 & 12.2  &       -     &       -     & 17.19(0.07) & TNT\\
04/06/02 & 52429.75 & 14.7  &       -     & 20.12(0.16) & 17.31(0.56) & KAIT\\
06/06/02 & 52431.75 & 16.7  &       -     & 19.72(0.12) & 17.24(0.46) & KAIT\\
08/06/02 & 52433.70 & 18.6  & $\geq20.46$ & 20.27(0.12) & 17.15(0.14) & N1mT\\
10/06/02 & 52436.37 & 21.3  & $\geq20.65$ &       -     & 17.20(0.02) & EKAR\\
11/06/02 & 52436.71 & 21.6  & $\geq20.65$ & 19.65(0.16) & 17.08(0.10) & N1mT\\
11/06/02 & 52436.75 & 21.7  &       -     & 19.62(0.20) & 17.25(0.24) & KAIT\\
14/06/02 & 52440.39 & 25.3  & $\geq20.58$ & 19.18(0.08) & 17.22(0.02) & EMMI\\
26/06/02 & 52452.40 & 37.3  &       -     & 20.09(0.08) & 18.16(0.05) & INT\\
27/06/02 & 52453.42 & 38.3  & $\geq21.39$ & 20.08(0.11) & 18.33(0.06) & INT\\
29/06/02 & 52455.43 & 40.3  &       -     &       -     & 17.85(0.30) & JKT\\
02/07/02 & 52458.45 & 43.4  &       -     & 20.07(0.23) & 17.82(0.07) & JKT\\
05/12/02 & 52613.93 & 198.8 & $\geq23.27$ & $\geq23.30$ &       -     & N1mT\\
12/12/02 & 52620.98 & 205.9 & $\geq23.85$ & $\geq23.41$ & $\geq23.28$ & N1mT\\
\hline
\end{tabular}
\begin{flushleft}
*Relative to $I_{max}$ (JD = 2,452,415.09).\\
N1mT = 1~m Nickel telescope + CCD, 0.28\arcsec pix$^{-1}$; TNT =
0.72~m Teramo-Normale Telescope + CCD, 0.46\arcsec pix$^{-1}$; KAIT
= 0.76~m Katzman Automatic Imaging Telescope + CCD, 0.80\arcsec pix$^{-1}$;
EKAR = 1.82~m Copernico telescope + AFOSC, 0.47\arcsec pix$^{-1}$; EMMI =
3.6~m ESO NTT + EMMI, 0.167\arcsec pix$^{-1}$; INT = 2.5~m Isaac Newton
Telescope + WFC, 0.33\arcsec pix$^{-1}$; JKT = 1.0~m Jacob Kapteyn Telescope
+ JAG, 0.33\arcsec pix$^{-1}$.\\
\end{flushleft}
\end{table*}

\begin{table*}
\centering
\caption{S-corrected optical photometry of SN 2002cv.}
\label{tabla_ori_ph_opt}
\begin{tabular}{lcrcccc}
\hline
   UT Date  & JD $-$    &Phase*&  $V$  &  $R$   &  $I$    & Instr.$^1$\\
         & 2,400,000.00 &(days)&               &               &               &       \\
\hline
14/05/02 & 52408.75 & $-$6.3 & $\geq20.97$  & 19.81(0.18) &       -     & N1mT\\
14/05/02 & 52409.37 & $-$5.7 & $\geq20.48$  &       -     & 17.23(0.14) & TNT\\
15/05/02 & 52409.74 & $-$5.4 & $\geq20.69$  & 19.62(0.06) & 17.45(0.02) & N1mT\\
15/05/02 & 52409.75 & $-$5.3 &       -      & 19.73(0.12) & 17.49(0.05) & KAIT\\
15/05/02 & 52410.39 & $-$4.7 &       -      &       -     & 17.18(0.12) & TNT\\
15/05/02 & 52410.40 & $-$4.7 &       -      &       -     & 17.32(0.04) & EKAR\\
16/05/02 & 52410.74 & $-$4.4 & $\geq20.50$  & 19.54(0.06) & 16.89(0.03) & N1mT\\
16/05/02 & 52411.41 & $-$3.7 &       -      & 19.42(0.10) & 16.69(0.02) & TNT\\
17/05/02 & 52411.75 & $-$3.3 &       -      & 19.28(0.09) &       -     & KAIT\\
17/05/02 & 52412.38 & $-$2.7 &       -      &       -     & 16.65(0.05) & TNT\\
20/05/02 & 52415.39 &  0.3 &       -      &       -     & 16.67(0.06) & TNT\\
23/05/02 & 52417.75 &  2.7 &       -      & 19.24(0.08) & 16.86(0.06) & KAIT\\
27/05/02 & 52421.75 &  6.7 &       -      & 19.28(0.12) & 17.06(0.08) & KAIT\\
28/05/02 & 52423.37 &  8.3 &       -      &       -     & 17.25(0.04) & TNT \\
29/05/02 & 52424.33 &  9.2 &       -      &       -     & 17.23(0.04) & TNT\\
30/05/02 & 52425.36 & 10.3 &       -      &       -     & 17.24(0.02) & TNT\\
31/05/02 & 52425.75 & 10.7 &       -      & 19.58(0.04) & 17.35(0.04) & KAIT\\
01/06/02 & 52427.28 & 12.2 &       -      &       -     & 17.19(0.02) & TNT\\
04/06/02 & 52429.75 & 14.7 &       -      & 20.05(0.16) & 17.33(0.28) & KAIT\\
06/06/02 & 52431.75 & 16.7 &       -      & 19.65(0.12) & 17.26(0.23) & KAIT\\
08/06/02 & 52433.70 & 18.6 & $\geq20.46$  & 20.27(0.12) & 17.15(0.03) & N1mT\\
10/06/02 & 52436.37 & 21.3 & $\geq20.72$  &       -     & 17.29(0.03) & EKAR\\
11/06/02 & 52436.71 & 21.6 & $\geq20.65$  & 19.65(0.07) & 17.08(0.03) & N1mT\\
11/06/02 & 52436.75 & 21.7 &       -      & 19.54(0.20) & 17.27(0.12) & KAIT\\
14/06/02 & 52440.39 & 25.3 & $\geq20.58$  & 19.26(0.08) & 17.21(0.02) & EMMI\\
26/06/02 & 52452.40 & 37.3 &       -      & 20.09(0.08) & 17.78(0.07) & INT\\
27/06/02 & 52453.42 & 38.3 & $\geq21.37$  & 20.07(0.03) & 17.96(0.06) & INT\\
29/06/02 & 52455.43 & 40.3 &       -      &       -     & 18.26(0.04) & JKT\\
02/07/02 & 52458.45 & 43.4 &       -      & 20.01(0.23) & 18.24(0.08) & JKT\\
05/12/02 & 52613.93 &198.8 & $\geq23.27$  & $\geq23.30$ &       -     & N1mT\\
12/12/02 & 52620.98 &205.9 & $\geq23.85$  & $\geq23.41$ & $\geq23.28$ & N1mT\\
\hline
\end{tabular}
\begin{flushleft}
*Relative to $I_{max}$ (JD = 2,452,415.09).\\
$^1$See note to Table \ref{tabla_noscor_ph_opt} for the telescope coding.\\
\end{flushleft}
\end{table*}

\begin{table*}
\centering
\caption{Original near-IR photometry of SN 2002cv.}
\label{tabla_ori_ph_ir}
\begin{tabular}{lcrcccc}
\hline
 UT Date    &  JD $-$   &Phase*&  $J$  &  $H$  &  $K$    &   Instr.\\
         & 2,400,000.00 &(days)&               &               &               &         \\
\hline
09/05/02$\triangleleft$ & 52404.39 & $-$10.7 & 17.05(0.11) &       -     & 16.64(0.11) & AZT\\
13/05/02$\triangleleft$ & 52408.35 & $-$6.7 & 15.46(0.01) & 14.92(0.02) & 14.73(0.05) & AZT\\
14/05/02$\triangleleft$ & 52409.35 & $-$5.7 & 15.26(0.01) & 14.61(0.02) & 14.36(0.03) & AZTDP\\
15/05/02$\triangleleft$ & 52410.32 & $-$4.8 & 15.07(0.02) & 14.56(0.03) & 14.25(0.04) & AZTDP\\
16/05/02$\triangleleft$ & 52411.36 & $-$3.7 & 14.90(0.01) & 14.44(0.02) & 14.21(0.03) & AZT\\
17/05/02$\triangleleft$ & 52412.38 & $-$2.7 & 14.89(0.01) & 14.34(0.01) &       -     & AZTDP\\
20/05/02$\triangleleft$ & 52415.33 &  0.2 & 14.75(0.03) &       -     & 13.95(0.03) & AZT\\
26/05/02$\triangleleft$ & 52421.43 &  6.3 &       -     &       -     & 14.06(0.04) & AZT\\
27/05/02$\triangleleft$ & 52422.36 &  7.3 &       -     &       -     & 14.06(0.05) & AZTDP\\
28/05/02$\triangleleft$ & 52423.41 &  8.3 & 15.37(0.04) &       -     &       -     & AZT\\
29/05/02$\triangleleft$ & 52424.38 &  9.3 &       -     & 14.74(0.02) &       -     & AZT\\
30/05/02 & 52425.36 & 10.3 &       -      &       -     & 14.16(0.03) & AZTDP\\
30/05/02 & 52425.42 & 10.3 & 15.69(0.07)  &       -     &       -     & AZT\\
31/05/02$\triangleleft$ & 52426.41 & 11.3 & 15.80(0.02) &       -     &       -     & AZTDP\\
01/06/02$\triangleleft$ & 52427.33 & 12.2 & 15.85(0.04) & 14.71(0.03) & 14.41(0.04) & AZTDP\\
05/06/02 & 52431.36 & 16.3 & 16.87(0.08)  & 14.49(0.04) & 14.94(0.07) & AZT\\
12/06/02 & 52438.35 & 23.3 & 16.33(0.03)  & 14.41(0.03) & 14.26(0.04) & AZT\\
16/06/02 & 52441.53 & 26.4 & 16.04(0.01)  &       -     & 14.17(0.01) & SofI\\
20/06/02 & 52446.36 & 31.3 & 15.77(0.05)  &       -     & 13.95(0.10) & AZT\\
04/07/02 & 52460.33 & 45.2 &       -      &       -     & 14.40(0.06) & AZT\\
09/07/02 & 52465.33 & 50.2 & 15.87(0.09)  &       -     &       -     & AZT\\
10/07/02 & 52466.33 & 51.2 &       -      & 15.39(0.10) &       -     & AZT\\
11/07/02 & 52467.32 & 52.2 &       -      &       -     & 14.89(0.07) & AZT\\
12/07/02 & 52468.33 & 53.2 & 16.50(0.10)  &       -     &       -     & AZT\\
22/07/02 & 52478.30 & 63.2 &       -      &       -     & 14.25(0.80) & AZT\\
\hline
\end{tabular}

\begin{flushleft}
*Relative to $I_{max}$ (JD = 2,452,415.09).\\
$\triangleleft$ Photometric night.\\
AZT = 1.08~m AZT-24 + SWIRCAM, 1.04\arcsec pix$^{-1}$; AZTDP =
data from \cite{dipaola02} using the same configuration as AZT;
SofI = 3.6~m ESO NTT + SofI, 0.29\arcsec pix$^{-1}$.\\
\end{flushleft}
\end{table*}

\subsection{Light curves} \label{ph_lc}

\begin{figure}
\begin{center}
\psfig{figure=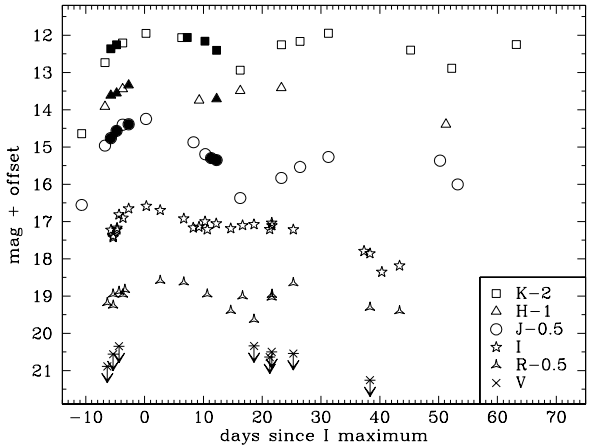,angle=270,width=9.5cm,height=7.1cm}
\caption{S-corrected $VRIJHK$ light curves of SN 2002cv during the
first weeks post-explosion. The original data measurements by Di
Paola et al. (2002) (AZTDP in Table \ref{tabla_ori_ph_ir}) are
marked as filled symbols. The light curves have been shifted by
the amount shown in the legend.} \label{fig_lightcurv}
\end{center}
\end{figure}

The early light curves are shown in Figure \ref{fig_lightcurv}.
Since SN~2002cv is not visible in the $B$ band, the phase is
relative to the epoch of the first $I$-band maximum, which occurred on
2002 May 20.6 (JD = 2,452,415.1 $\pm$ 0.2).

The light curves have good sampling from $-10$ to +64 days after $I$
maximum, except in $V$ where only upper limits could be measured.
SN~2002cv was discovered in the $JK$ bands about 10.7 days before
the $I$ maximum, making these measurements among the earliest NIR
observations available for a SN~Ia.

A secondary maximum in the red and NIR light curves is easily
visible, which is typical of SNe~Ia and confirms the early
classification of SN~2002cv. We remind the reader that no definite
classification was assigned to SN~2002cv because of the lack of
clear spectral features blueward of 6000~\AA. Another typical SN~Ia
signature is the first $I$ maximum occurring before those in $R$
and the NIR (e.g., \citealt{contardo00}).

The $R$-band light curve presents a pronounced secondary maximum,
which is unusual for a SN~Ia. This is probably due to the high
reddening suffered by SN~2002cv, which shifts the effective
wavelength of the $R$ bandpass to the red, mimicking the $I$ light
curve of an unreddened SN~Ia. In fact, we verified that given the
SN~2002cv spectrum, the actual effective wavelength shifts from
$\sim$6550~\AA\ to $\sim$7100~\AA, which leads to a difference in
magnitude up to 0.3, mainly visible at the phase of the secondary
maximum. This effect is negligible in the $IJHK$ bands.

We note also that the post-maximum decline of the $J$ light curve is
much steeper than that in the $I$ band, showing a pronounced $J$
minimum around day +16. On the other hand, a shallow $H$ minimum
occurs a few days earlier than those in $J$ and $K$. All these
features, except the broad $K$ peak, are typical of SNe~Ia
\citep{meikle00}.

\begin{figure}
\begin{center}
\psfig{figure=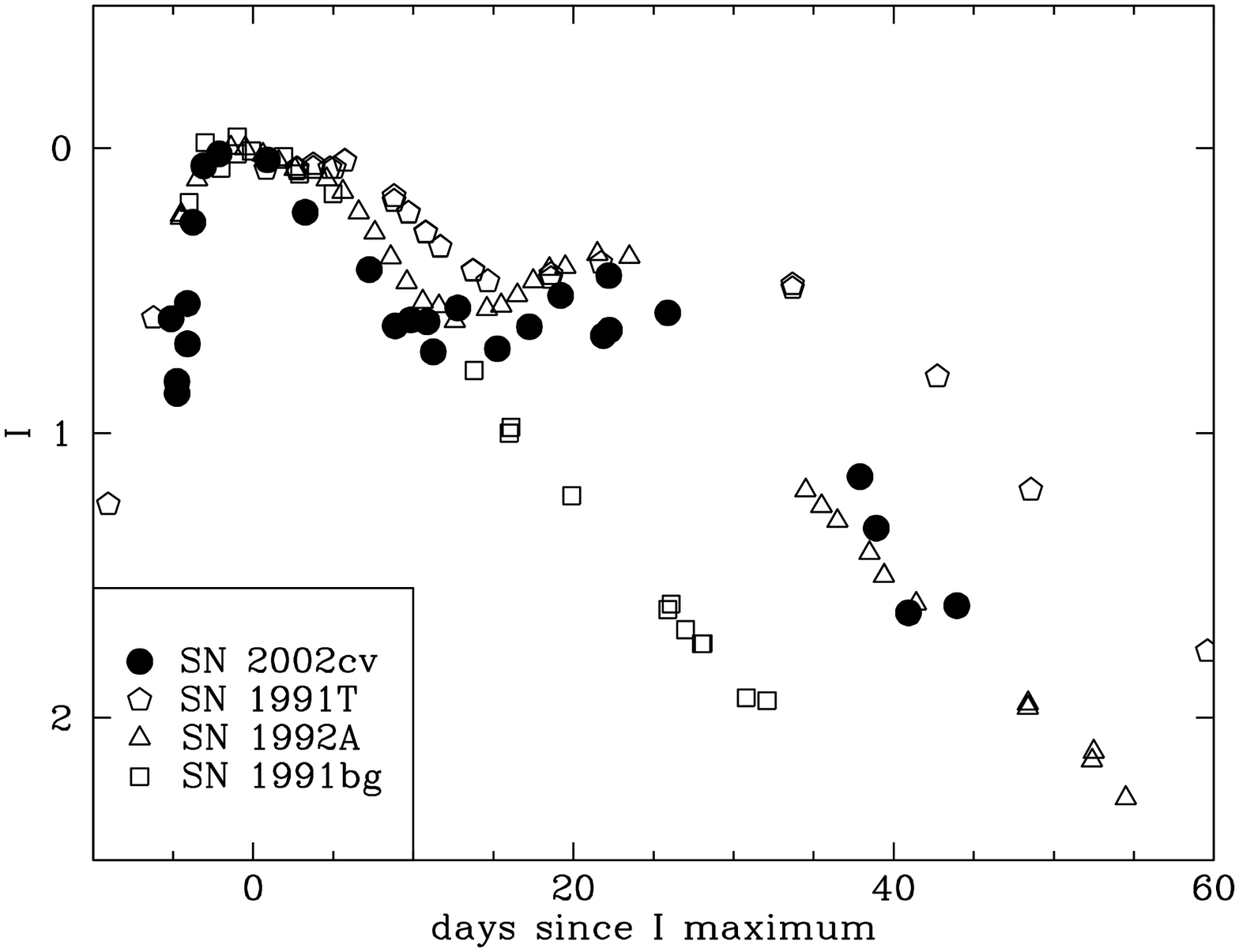,angle=270,width=9.5cm,height=7.1cm}
\caption{Comparison between the $I$-band light curve of SN~2002cv
and three other SNe~Ia with different \dm15 values: SN~1991T (\dm15
= 0.94 mag), SN~1991bg (\dm15 = 1.94 mag), and SN 1992A (\dm15 =
1.47 mag) normalized to the $I$ maximum. See Table \ref{tabla_mainsne}
for more information about these SNe.}\label{fig_lcdm15}
\end{center}
\end{figure}

For comparison, in Figure \ref{fig_lcdm15} we plot the $I$ light
curves of three nearby SNe~Ia (see Section \ref{param} for
more information) having different values of \dm15: SN~1991T
(\dm15 = 0.94 mag), SN~1991bg (\dm15 = 1.94 mag), and SN~1992A
(\dm15 = 1.47 mag). The light
curves have been shifted to match the first $I$ maximum. The light
curve of SN~2002cv is best matched by that of SN~1992A even if in
the former the first maximum is slightly narrower. Assuming that
the two SNe were also similar in the $B$ band, we can assign
tentatively to SN~2002cv a value \dm15 $\approx$ 1.5 mag. We
obtain a better estimate of \dm15 in Section \ref{param_value}.\\

\subsection{Colour and pseudo bolometric curves} \label{ph_cc}

\begin{figure}
\centering \psfig{figure=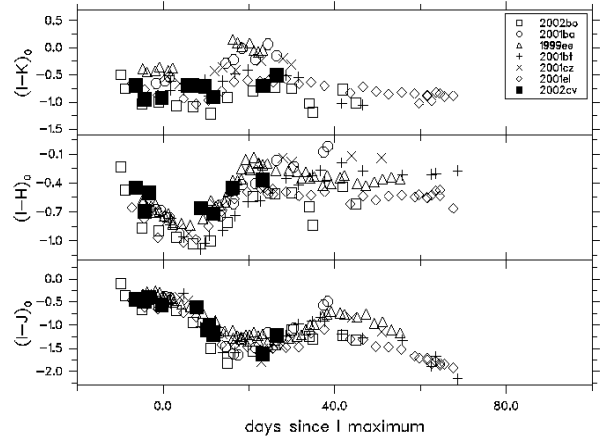,width=9cm,height=6.5cm}
\caption{Colour evolution of SN~2002cv compared with those of SNe
2002bo (\citealt{benetti04}; \citealt{krisc04b}), 2001bt, 2001cz
(\citealt{krisc04b}), 1999ee, 2001ba (\citealt{krisc04a}), and
2001el (\citealt{krisc03}). The best fit is obtained for a
reddening $A_V$ = 8.99 $\pm$ 0.30 mag and total-to-selective
extinction ratio $R_V$ = 1.97 $\pm$ 0.30.} \label{fig_color}
\end{figure}

In Figure \ref{fig_color} the evolution of the intrinsic ($I$-NIR)
colours for SN~2002cv (corrected for the reddening as discussed in
Section \ref{redd}) are compared with those of a sample of SNe~Ia.
The colour curves are in general very similar to those of normal
SNe~Ia such as SN~2001cz \citep{krisc04b} or SN~2001el
\citep{krisc03}, but with some differences in $(I-K)_0$.

\begin{figure}
\centering \psfig{figure=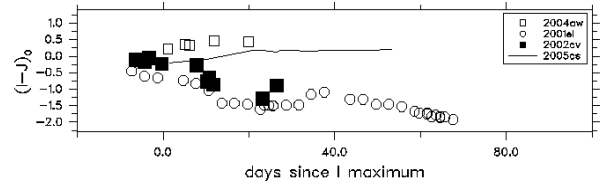,width=9cm,height=3cm}
\caption{Colour evolution of SN~2002cv compared with that of
SN~2001el (Type Ia, \citealt{krisc03}), SN~2004aw (Type Ic,
\citealt{taubenberger06}), and SN~2005cs (Type IIP,
\citealt{pastorello06}; \citealt{pastorello07}). The curves have
been dereddened according to the values reported in the mentioned
papers.} \label{fig_colorII}
\end{figure}

The good match of the SN~2002cv colour curve with those of other
normal SNe~Ia is another confirmation of the Type Ia
classification of this SN. This is strengthened by the comparison
presented in Figure \ref{fig_colorII}, in which the $(I-J)_0$
curve of SN~2002cv is compared with those of different types of SNe:
SN~2001el (Type Ia, \citealt{krisc03}), SN~2004aw (Type Ic,
\citealt{taubenberger06}), and SN~2005cs (Type IIP,
\citealt{pastorello06}; \citealt{pastorello07}). During the
pre-maximum period all curves have similar colours, but thereafter
the three SN types follow different patterns, with SN~2002cv
turning to the blue, as did SN~Ia 2001el.

\begin{figure}
\centering
\psfig{figure=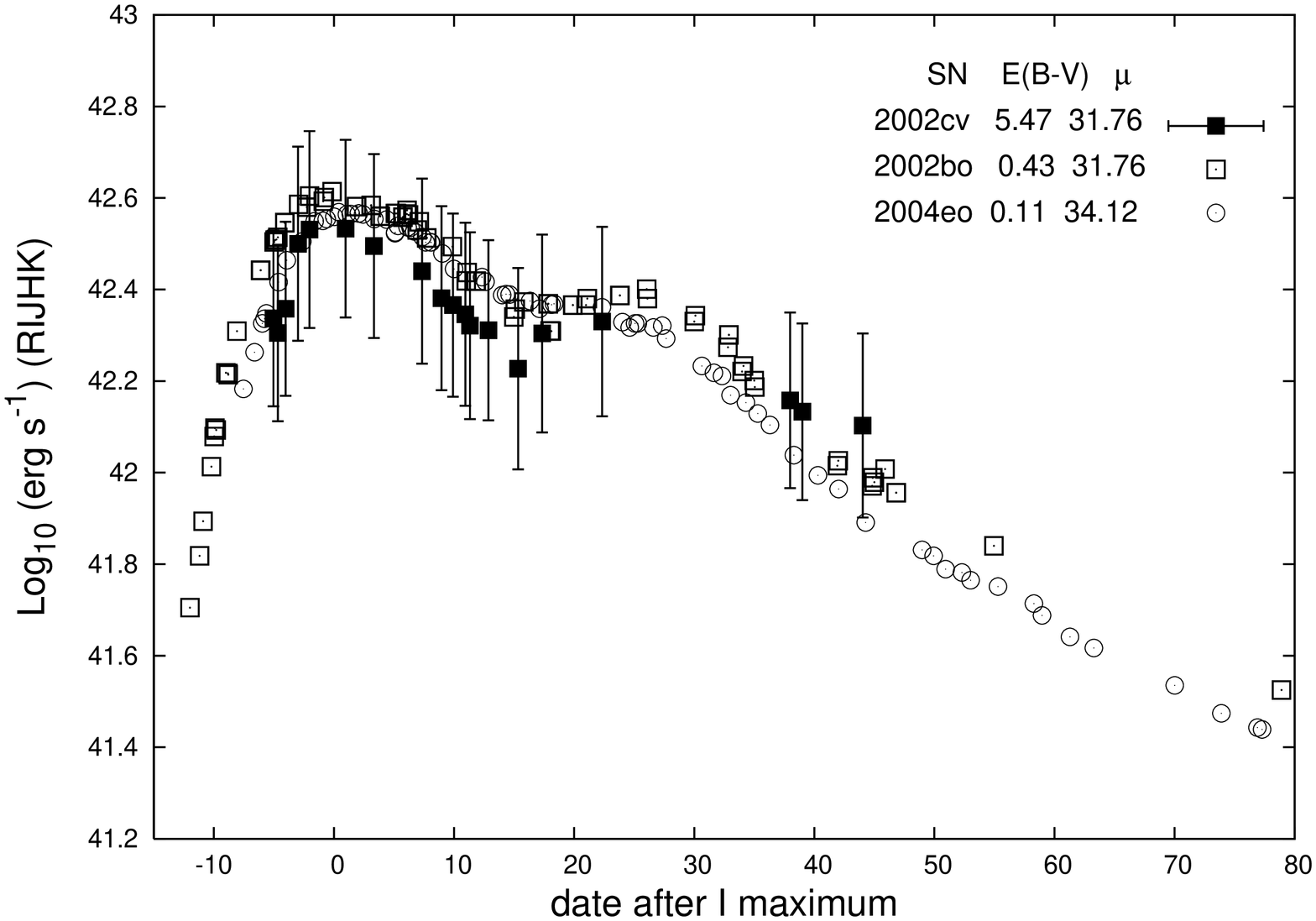,width=8.2cm,height=5.7cm,angle=270}
\caption{Pseudo-bolometric ($RIJHK$) light curve for SN~2002cv
(filled squares). Open squares and circles give the
pseudo-bolometric ($RIJHK$) light curves for SN~2002bo and
SN~2004eo, respectively. Error bars include photometric, reddening,
and distance uncertainties.} \label{fig_bolometric}
\end{figure}

Figure \ref{fig_bolometric} shows the ``pseudo-bolometric''
luminosity evolution of SN~2002cv derived by integrating the flux
in the $RIJHK$\footnote{We have checked that changing the effective
wavelength of the $R$ band in the calculation of the bolometric light
curve results in differences of 0.1 dex in the luminosity. This
was taken into account in the error estimate.} bands (with the
distance modulus and reddening discussed in Section \ref{redd}).
Total uncertainties were computed taking into account photometry errors,
and the uncertainties in reddening and distance. We have also
plotted the pseudo-bolometric light curves for SN~2002bo and
SN~2004eo. The reddening-corrected bolometric luminosity at
maximum is log$_{10}L$ ($RIJHK$) = 42.54 $\pm$ 0.30 \ergs. While the
first maximum of SN~2002cv is fainter and narrower than those of
the other SNe~Ia, its secondary maximum is bright and broad, similar
to that of SN~2004eo. This is not unexpected, since an overall
similarity between SN~2004eo and SN~1992A was noted by
\cite{pastorello06b}.

It is well known that for SNe~Ia around maximum, most of the flux is
emitted in the $UBV$ bands. From the photometry of other SNe~Ia,
we can estimate that the integrated flux carried by the $RIJHK$
bands is approximately 40\% (with a variation around 10\%) of the
integrated ``{\it uvoir}'' flux from the $U$ through $K$ bands
(see also Figure 6 of \citealt{contardo00}). Therefore, we estimate
that the total {\it uvoir} luminosity at maximum of SN~2002cv was
log$_{10}L_{{\it uvoir}}$ = 42.94 $\pm$ 0.60 \ergs.\\

Considering that in general, the $B$ maximum occurs approximately
three days after the $I$ maximum, we can derive the epoch of the $I$
secondary maximum relative to $B$ maximum and read the $^{56}$Ni
mass from Figure 11 of \cite{kasen06}. In that work, the prominence
and timing of the $I$ and $J$ secondary maxima were measured and plotted
vs. M$_{Ni}$ for $I$ and $J$-band models. \cite{kasen06} also computed
the $^{56}$Ni mass of their models considering the absolute maximum and
the deep minimum in the $J$ band (Figure 10 of \citealt{kasen06})
and reproduced synthetic model light curves by varying the mass of
$^{56}$Ni. These models suggest that SN~2002cv has $M$($^{56}$Ni)
$\approx$ 0.45 \M. This is in good agreement with the $^{56}$Ni mass
derived from the total {\it uvoir} luminosity of SN~2002cv at
maximum, the Arnett rule \citep{arnett82,stritz05}, and a rise
time of 19 days, which gives an estimate of the $^{56}$Ni mass
$M$($^{56}$Ni) = 0.44 $\pm$ 0.07 \M. This is also remarkably similar
to the $^{56}$Ni mass derived by \cite{pastorello06b} for
SN~2004eo (0.45 \M).

\section{Spectroscopy} \label{spec}

\subsection{Data reduction} \label{spec_obs}

The spectroscopic observations are summarised in Table
\ref{tabla_spec} and the instruments used are as follows:

\begin{enumerate}

\item 1.82~m Copernico telescope of Mt. Ekar (Asiago, Italy),
equipped with AFOSC (see Section \ref{ph_obs}). Grism\#2
(wavelength region 5250--10,300~\AA, dispersion 15.67~\AA\
pix$^{-1}$, and resolution 38~\AA\ with the 2.1\arcsec slit)
was used;

\item 3~m Shane reflector (Lick Observatory, Mt. Hamilton,
California, USA), equipped with Kast (double spectrograph with
simultaneous red/blue spectra, wavelength range 3300--10,400~\AA);

\item 3.6~m ESO/NTT telescope (La Silla, Chile), equipped with EMMI
(see Section \ref{ph_obs}). For low-dispersion spectroscopy,
grism\#2 (wavelength range 3800--9700~\AA, dispersion 1.74~\AA\
pix$^{-1}$) was used;

\item 3.8~m United Kingdom Infrared Telescope (Mauna Kea, Hawaii,
USA), equipped with CGS4 (Cooled Grating Spectrometer Mk 4, 1.22
\arcsec pix$^{-1}$). The 40 line mm$^{-1}$ grating and the $IJHK$
filters (wavelength range 8250--25,100~\AA) were used;

\item 3.6~m ESO/NTT telescope (La Silla, Chile), equipped with SofI
(see Section \ref{ph_obs}). The grism GBF (wavelength range
9450-16,520~\AA, dispersion 6.96~\AA\ pix$^{-1}$, and resolution
$\sim$30~\AA) was used.

\end{enumerate}

The spectra were reduced using IRAF and FIGARO (for NIR data)
routines. The pre-processing of the spectroscopic data (trimming,
bias, overscan, and flat-field correction) were the same as for the
imaging. For the NIR spectra, before extraction, the contribution of
the night-sky lines was removed from the two-dimensional NIR spectrum,
subtracting another two-dimensional spectrum with the target placed
in a different position along the slit. The one-dimensional
spectra were wavelength calibrated by comparison with arc-lamp
spectra obtained during the same night and with the same
instrumental configuration, and flux calibrated using
spectrophotometric standard stars. The zero-point of the
wavelength calibration was verified against the bright night-sky
emission lines. The standard-star spectra were also used to model
and remove the telluric absorption. The absolute flux calibration
of the spectra was checked against the photometry and when
necessary, the spectra were re-scaled. After that, the typical
deviation from photometry is less than 10\% in all bands.

\begin{table*}
\centering
\caption{Optical and NIR spectroscopic observations of SN 2002cv.}
\label{tabla_spec}
\begin{tabular}{lcrccc}
\hline
 UT Date    &  JD $-$  &Phase*&Grism/Grating& Range     &Instr.\\
         & 2,400,000.00 &(days)&             & (\AA)   &      \\
\hline
15/05/02 & 52410.42 & $-$4.7 & gm2       & 5250--10,300 & Ekar \\
19/05/02 & 52413.77 & $-$1.3 & $\ddagger$& 3300--10,400 & Shane \\
08/06/02 & 52433.75 & +18.7& $\ddagger$& 3300--10,400 & Shane \\
15/06/02 & 52440.52 & +25.4& gm2       & 3900--9700 & EMMI \\
\hline
\hline
 UT Date    & JD $-$   &Phase*&Grism/Grating&   Range    &Instr.\\
         & 2,400,000.00 &(days)&             &  (\AA)   &      \\
\hline
22/05/02 & 52417.25 & +2.2 & gmij      & 8250--13,500 & UKIRT \\
23/05/02 & 52418.25 & +3.2 & gmk       & 19800--25,100 & UKIRT \\
15/06/02 & 52441.49 & +26.4& gmB       & 9450--16,512 & SofI \\
\hline
\end{tabular}
\begin{flushleft}
*Relative to $I_{max}$ (JD = 2,452,415.09).\\
$\ddagger$ grating 300/7500 + grism 600/4310.\\
Ekar = 1.82~m Copernico Telescope + AFOSC; Shane = 3~m Shane
reflector + Kast dual spectrograph; EMMI = 3.6~m ESO NTT +
EMMI; UKIRT = 3.8~m United Kingdom Infrared Telescope + CGS4;
SofI = 3.6~m ESO NTT + SofI.\\
\end{flushleft}
\end{table*}

\subsection{Optical and NIR spectra} \label{spec_ev}

SN~2002cv was a faint target, especially in the optical: we were
able to secure only four spectra in the optical
and three in the NIR (Table \ref{tabla_spec}).\\

\begin{figure}
\begin{center}
\psfig{figure=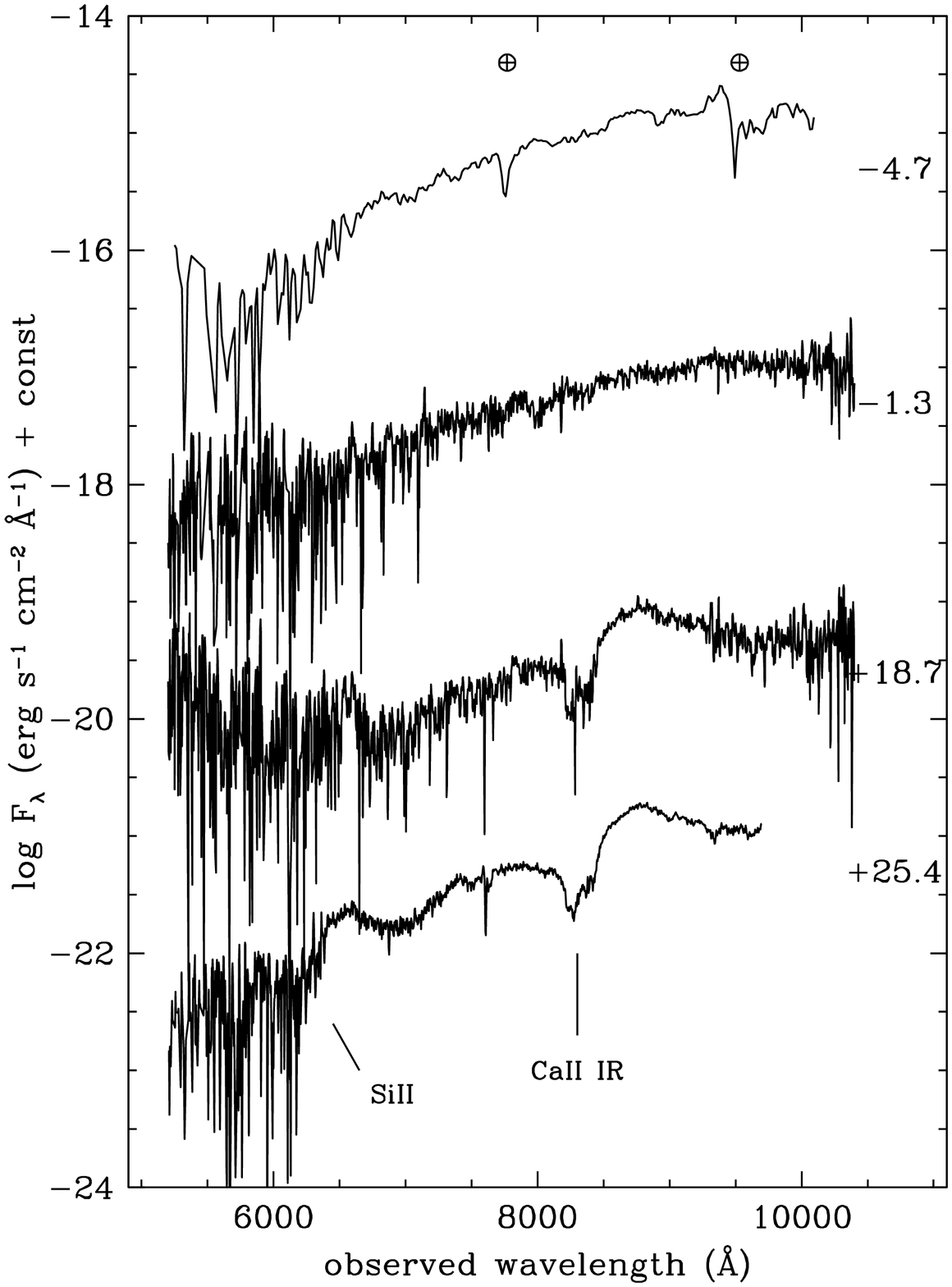,width=8cm,height=10.6cm}
\caption{Optical spectra of SN~2002cv, not corrected for
reddening. The ordinate refers to the first spectrum and the
others have been shifted downward by arbitrary amounts (0, $-1$,
$-3$, and $-5.5$ from top to the bottom). Epochs (days) relative
to $I$ max are given at the right-hand side.}
\label{fig_spec_optevol}
\end{center}
\end{figure}

The sequence of optical spectra of SN~2002cv is shown in Figure
\ref{fig_spec_optevol}. The spectra are distributed from $-4.7$ days
to +25.4 days relative to the $I$-band maximum. Spectra are truncated
below 5000~\AA\ because at shorter wavelengths no significant signal
from the SN was detected. The most notable feature of the optical
spectra is the very red continuum, with almost no sign of
individual lines except for the \CaII~NIR triplet ($\sim$ 8500 \AA),
which is clearly seen two weeks after maximum. In the last spectra
there might also be evidence of the \SiII~absorption at $\sim$6350~\AA.

\begin{figure}
\begin{center}
\psfig{figure=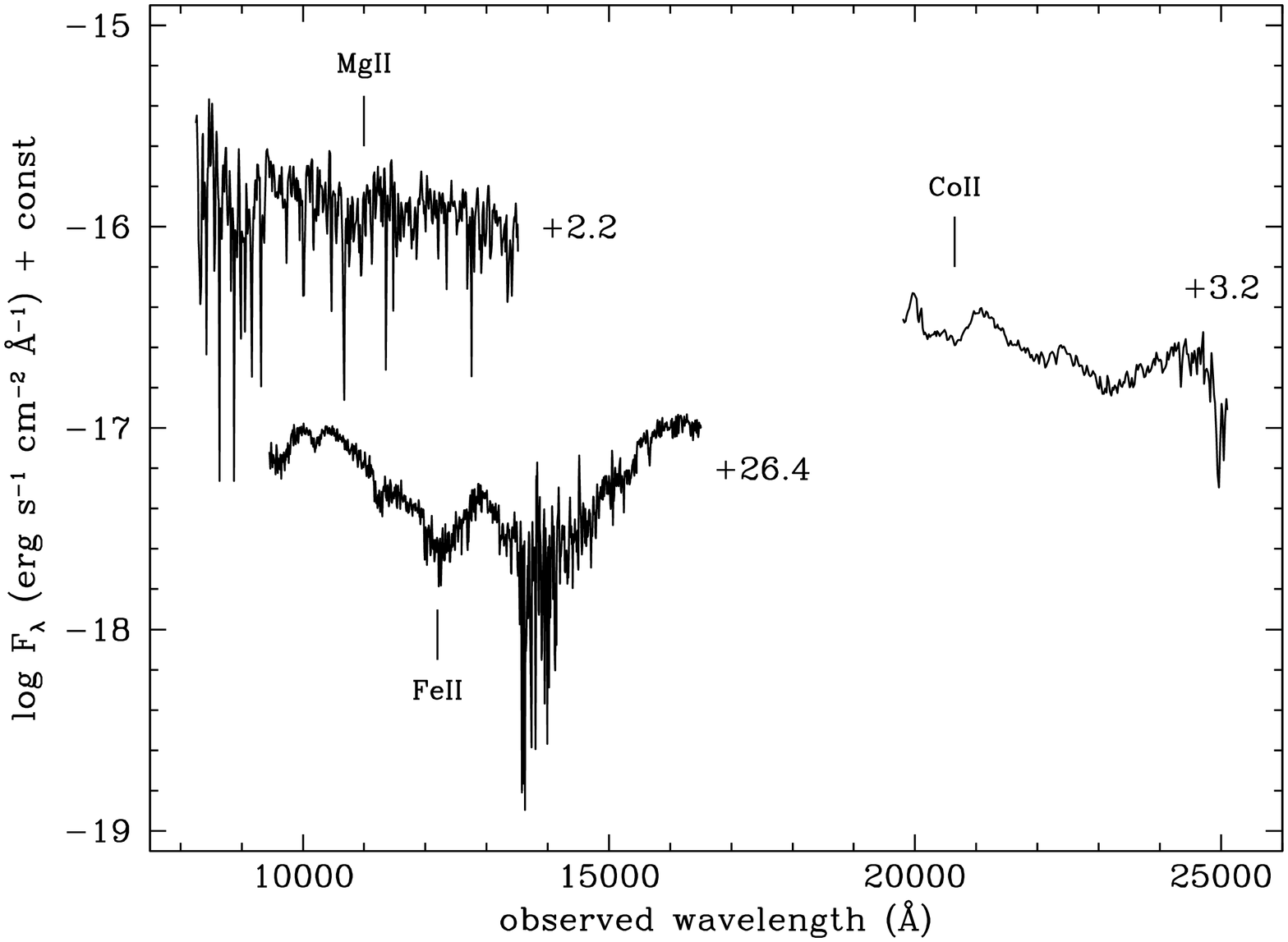,angle=270,width=8.8cm,height=6.6cm}
\caption{NIR spectra of SN~2002cv, not corrected for reddening.
The ordinate refers to the first spectrum and the others have been
shifted downward by arbitrary amounts (0, $-1$, and $-2.5$ from
the top to the bottom). Epochs are given at the right-hand side.}
\label{fig_spec_irevol}
\end{center}
\end{figure}

The NIR spectra of SN~2002cv are shown in Figure
\ref{fig_spec_irevol}. The spectral evolution is typical of
SNe~Ia. The earliest two spectra are dominated by the continuum.
The spectrum at phase +2.2 days shows a weak feature with P-Cygni
profile at about 10,900~\AA\ (rest wavelength) possibly due to
\MgII\ \citep{wheeler98}. Weak emission is also visible at
20,500~\AA\ on day +3.2, perhaps due to \CoII\ lines
\citep{marion03}. In the +26.4 day spectrum, several individual
broad features are remarkably strong. In particular a prominent
absorption at $\sim$12,300~\AA\ \citep{marion03} is attributed to
\FeII. At about 15,000~\AA, SN~2002cv shows the rapid flux
turnover also observed in other SNe~Ia. According to
\cite{wheeler98} and \cite{marion03}, this occurs because the
region between 11,000 and 15,000~\AA\ has fewer blends of
iron-group lines compared to adjacent wavelengths.

\begin{figure}
\begin{center}
\psfig{figure=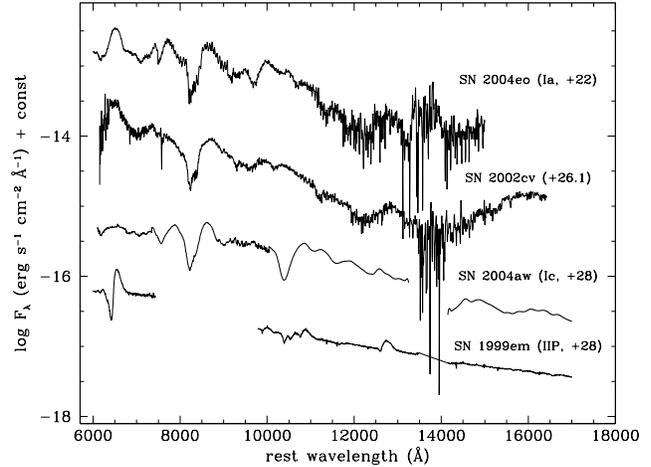,angle=270,width=8.8cm,height=6.6cm}
\caption{Comparison between the combined optical and NIR spectra
of SN~2002cv at +25.4 and +26.4 days after $I$ maximum
(respectively) with those of SN~2004eo (SN~Ia,
\citealt{pastorello06b}), SN~2004aw (SN~Ic,
\citealt{taubenberger06}), and SN~1999em (SN~IIP,
\citealt{hamuy01,leonard02}). All of the spectra have been
corrected for redshift and reddening.}\label{fig_spec_confr}
\end{center}
\end{figure}

Figure \ref{fig_spec_confr} shows the combined, nearly coeval
optical and NIR spectra of SN~2002cv (NTT+EMMI at +25.4 days and
NTT+SofI at phase +26.4 days) compared with similar-age spectra of
SN~2004eo (SN~Ia, \citealt{pastorello06b}), SN~2004aw (SN~Ic,
\citealt{taubenberger06}), and SN~1999em\footnote{The
SN~1999em spectra were downloaded from SUSPECT (The Online
Supernova Spectrum Archive):
http://bruford.nhn.ou.edu/~suspect/index1.html .} (SN~IIP,
\citealt{hamuy01,leonard02}). This comparison definitely confirms
that SN~2002cv is a SN~Ia (see, in particular, the overall
similarity to SN~2004eo). Moreover, the pronounced P-Cygni
profile of H$\alpha$ and P$\beta$ of SNe~II (SN~1999em,
6562.8~\AA\ and 12,818~\AA), and the \HeI~and \CI~around 10,830~\AA\
of SNe~Ic (SN~2004aw), are not present in SN~2002cv.

\section{The Reddening Estimate} \label{redd}

We mentioned in the introduction that extinction can be the
dominant source of error in the spectrophotometric calibration of
SNe~Ia. This is especially true for SN~2002cv because of
the very high extinction.

\cite{dipaola02} derived an estimate of the extinction of
SN~2002cv after assuming typical $J$, $H$, and $K$ absolute
magnitudes for SNe~Ia, as follows:

\begin{equation}
A_{\lambda} = m_{\lambda} - M_{\lambda} - \mu. \label{equ_magabs}
\end{equation}

\noindent
Using a standard extinction law, these NIR estimates were converted
to an average \av\ = 7.90 $\pm$ 0.90 mag. However, we found that using
this approach for different bands gives values of \av\ which are
inconsistent, with progressively lower values moving from red to
blue bands. As in other similar cases (e.g., \citealt{krisc06a},
\citealt{elias06}), the simplest explanation is that \rv\ has a
value smaller than the standard 3.1.

To derive consistent estimates of both \av\ and \rv, we adopt
three different procedures, all based on the comparison of the SED
and the luminosity of SN~2002cv with those of other standard
SNe~Ia.

\begin{enumerate}

\item We performed a simultaneous match of the $I-J$, $I-H$, and $I-K$
colour curves of SN~2002cv with those of other normal SNe~Ia
(Figure \ref{fig_color}). As in the case of SN~2003cg
\citep{elias06}, it turns out that the best fit requires a value
of \rv\ smaller than the canonical one. The best match is for \av\
= 8.99 $\pm$ 0.30 mag and \rv\ = 1.97 $\pm$ 0.30.

It is interesting to apply the same methods to SN~2002bo, which
exploded in the same host galaxy as SN~2002cv. In this case we can
make use of more colours, namely $B-V$, $V-R$, $R-I$, $V-J$, $V-H$, and
$V-K$. Matching the colours of SN~2002bo to those of normal SNe~Ia
(Figure \ref{fig_color02bo}) gives a best fit of \av\ = 1.50 $\pm$
0.30 mag and \rv\ = 2.99 $\pm$ 0.30. This dust-extinction
law is consistent with the average one in the Galaxy. However, we
note in Figure \ref{fig_color02bo} that the $V$-NIR colour curves of
SN~2002bo do not fit those of normal SNe~Ia well. This was already
noted by \cite{krisc04a} and attributed to a real peculiarity in
the SED of this SN.

Since \rv\ is related to the grain size, the different extinction
laws found for SN~2002bo and SN~2002cv show that in NGC~3190, dust
having different average properties coexists. This is an important
fact but at the same time not surprising, since in the Galaxy we
also observe regions with very different values of \rv\
\citep{fitzpatrick04,geminale05}.

\begin{figure}
\begin{center}
\psfig{figure=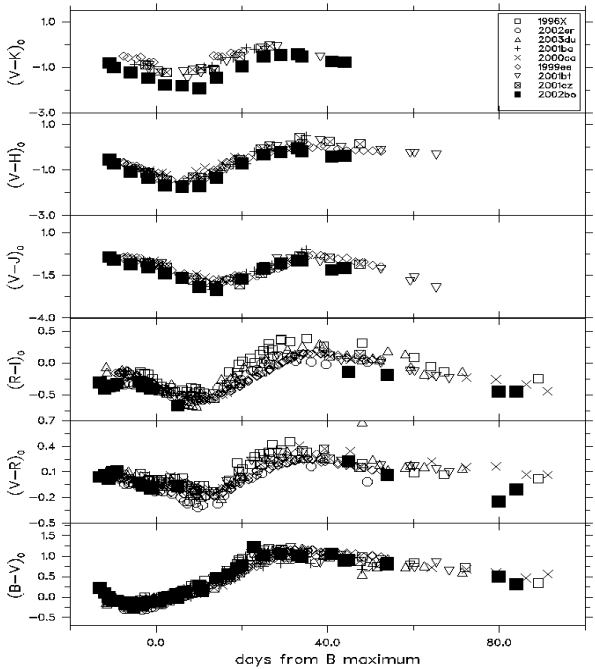,width=9cm,height=11.1cm}
\caption{As in Figure \ref{fig_color} for SN~2002bo. The best
match of the curves is obtained for \av\ = 1.50 $\pm$ 0.30 mag and
\rv\ = 2.99 $\pm$ 0.30.} \label{fig_color02bo}
\end{center}
\end{figure}

\item The observed extinction curve is derived by comparing the
observed optical ($>6000$~\AA) and NIR SED of SN~2002cv with those
of unreddened SNe~Ia, at similar epochs. The spectra used for this
comparison were previously corrected for redshift and Galactic
reddening, and scaled to the distance of SN~2002cv (an example is
shown in Figure \ref{fig_redspec}). This method was applied by
\cite{elias06} to derive the extinction law to SN~2003cg.

We used as templates the spectra of SN~1992A, SN~1994D, SN~1996X,
and SN~2004eo (see Section \ref{param} for more details about
these SNe), comparing a total of twelve pairs of spectra at
different epochs and obtaining in each case an estimate of \av\
and \rv. The average values are $A_{V,host}$ = 8.17 $\pm$ 0.57 mag
and \rv\ = 1.52 $\pm$ 0.11.

\begin{figure}
\begin{center}
\psfig{figure=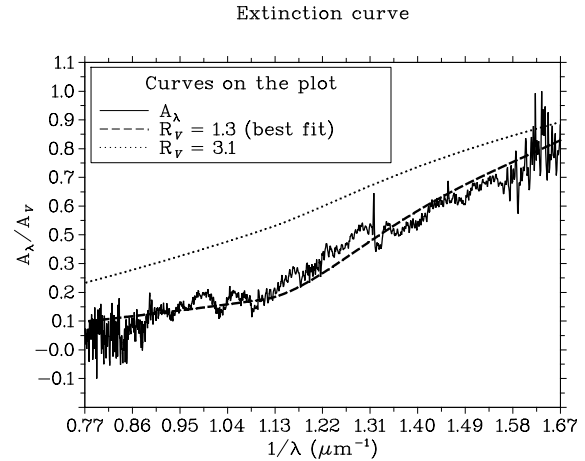,angle=270,width=9.5cm,height=6.9cm}
\caption{Best fit of the observed extinction law
($A_{\lambda}/A_V$) with the theoretical CCM laws (dashed line).
In the example, the extinction curve of SN~2002cv was obtained by
dividing the spectrum of SN~2002cv on day +26 by that of SN~2004eo
from 6000 to 12,500~\AA. For comparison, we also plot the CCM
extinction curve for \rv\ = 3.1 (dotted line). From the average of
12 comparisons using 4 different SNe as templates (see the text
for more information), we obtained $A_{V,host}$ = 8.17 $\pm$ 0.57
mag and \rv\ = 1.52 $\pm$ 0.11.} \label{fig_redspec}
\end{center}
\end{figure}

\item Exploiting the fact that SN~2002bo exploded in the same
galaxy as SN~2002cv, we performed a multi-dimensional
maximum-likelihood estimate\footnote{The maximum-likelihood method
is the procedure of finding the value of one or more parameters for
a given statistic which makes the known likelihood distribution a
maximum \citep{myung03}.} to find the values of \av, \rv, and
$\mu$ that give the best match to the measurements in columns 3
and 5 of Table \ref{tabla_reddcont} according to the relation

\begin{equation}
M^{'}_{\lambda}(A_V, R_V, \mu) = m_{\lambda} - \mu -
A_{\lambda,Gal} - [A_V(a_{\lambda} +
\frac{b_{\lambda}}{R_V})]_{host} \label{equ_magabs},
\end{equation}

\noindent
where $a_{\lambda}$ and $b_{\lambda}$ are wavelength-dependent
coefficients given by CCM. Here we remind the reader of the shift
in the $R$ effective wavelength due to the high extinction of SN~2002cv
(see Section \ref{ph_lc}). This effect is smaller at maximum light than
in the later phases, and hence we included the $R$-band measurements
in our fit (Table \ref{tabla_reddcont}). For a simultaneous fit to
the data of the two SNe, the free parameters are $A_{V,02cv}$,
$A_{V,02bo}$, $R_{V,02cv}$, $R_{V,02bo}$, and $\mu$ = $\mu_{02cv}$
= $\mu_{02bo}$.

Actually, following the results of method (i), we fixed for
SN~2002bo a standard value \rv\ = 3.1. Also, due to the different
$V$-NIR colour curves of SN~2002bo (see \citealt{krisc04a} and
method (i)), we did not include the NIR bands of this SN in our
fit.

Figure \ref{fig_contours} shows the $1\sigma$, $2\sigma$, and $3\sigma$
confidence levels projected in the planes of selected parameter pairs.
The uncertainty in \rv\ and \av\ can be read from the 1$\sigma$
contour. The maximum-likelihood test gives the following results:
\av\ = 8.40 $\pm$ 0.35 mag, \rv\ = 1.60 $\pm$ 0.10 for SN~2002cv,
\av\ = 1.00 $\pm$ 0.10 mag for SN~2002bo, and a distance modulus
of $\mu$ = 31.76 $\pm$ 0.07 mag for NGC 3190.

\end{enumerate}

\begin{table}
\centering \caption{Basic input data to find the values of \av,
\rv, and $\mu$ by maximum-likelihood
estimation.}\label{tabla_reddcont}
\setlength\tabcolsep{2pt}
\begin{tabular}{lccccc}
\hline
 Filter & $m^{02cv}_{\lambda,max}$$^1$ &  $M^{02cv}_{\lambda,max}$$^2$ & $m^{02bo}_{\lambda,max}$$^3$ &  $M^{02bo}_{\lambda,max}$$^4$ &  $A_{\lambda,Gal}$ \\
\hline
$B$  &  --  &  --  & $14.04 \pm 0.10$ & $-19.27 \pm 0.04$ & 0.102 \\
$V$  &  --  &  --  & $13.58 \pm 0.10$ & $-19.20 \pm 0.04$ & 0.078 \\
$R$  &  $19.08 \pm 0.20$ & $-19.04 \pm 0.13$ & $13.49 \pm 0.10$ & $-19.21 \pm 0.04$ & 0.063 \\
$I$  &  $16.57 \pm 0.10$ & $-18.79 \pm 0.12$ & $13.52 \pm 0.10$ & $-18.94 \pm 0.04$ & 0.046 \\
$J$  &  $14.75 \pm 0.03$ & $-18.61 \pm 0.13$ & --  &  -- & 0.022 \\
$H$  &  $14.34 \pm 0.01$ & $-18.28 \pm 0.15$ & --  &  -- & 0.015 \\
$K$  &  $13.91 \pm 0.04$ & $-18.44 \pm 0.14$ & --  &  -- & 0.009 \\
\hline
\end{tabular}
\begin{flushleft}
$^1$Observed magnitudes at maximum of SN~2002cv for each band
(see Section \ref{param_value}). \\
$^2$Adopted absolute magnitudes of SN~2002cv. For $R$ and $I$ see
\ref{param_value}; for $J$, $H$, and $K$ we used the mean absolute
magnitudes given by \cite{krisc04b}.\\
$^3$Observed magnitudes at maximum of SN~2002bo
\citep{benetti04}. \\
$^4$Adopted absolute magnitudes of SN~2002bo \citep{prieto06}.\\
\end{flushleft}
\end{table}

\begin{figure}
\begin{center}
\psfig{figure=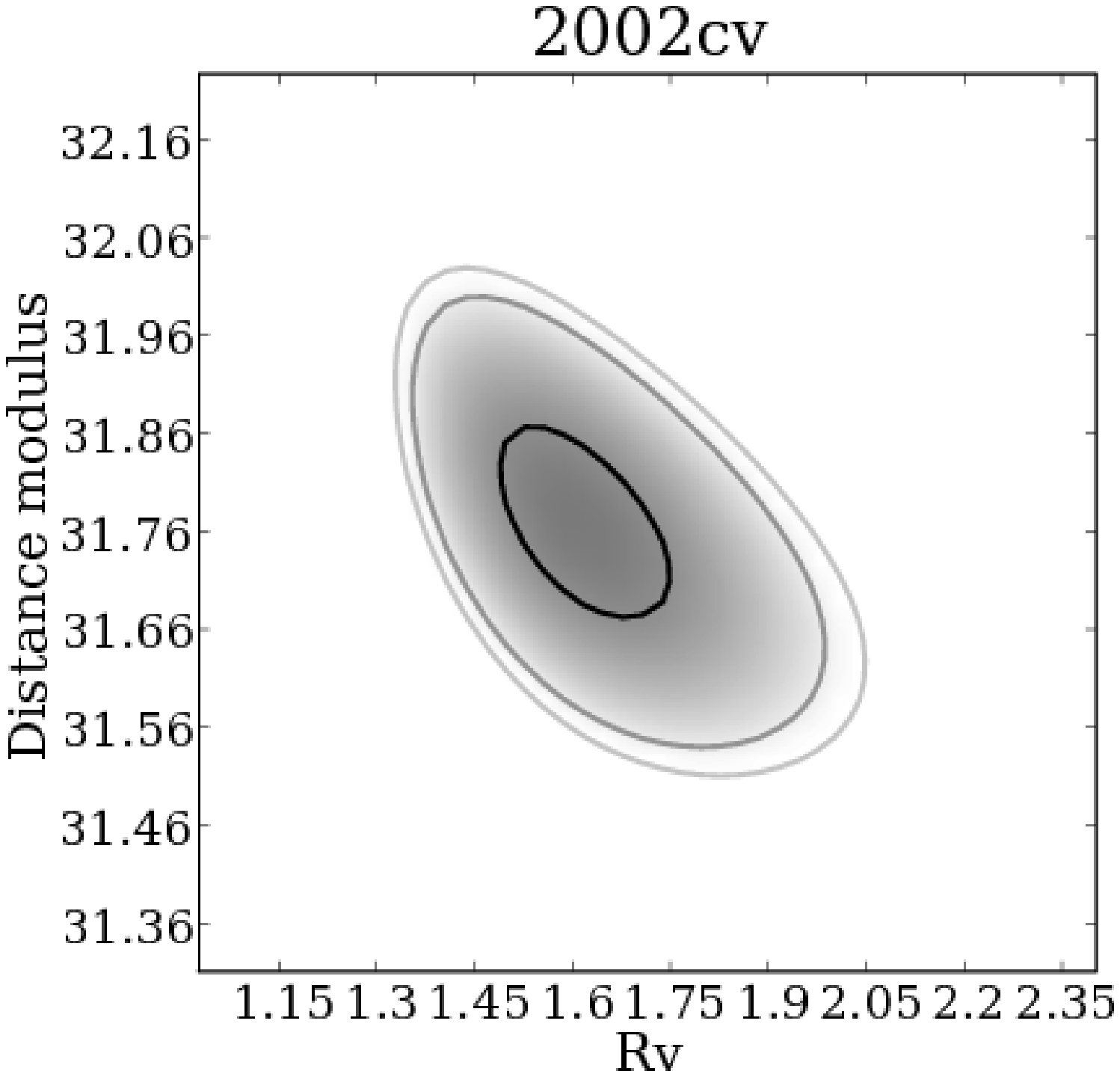,width=9cm,height=5.5cm}
\psfig{figure=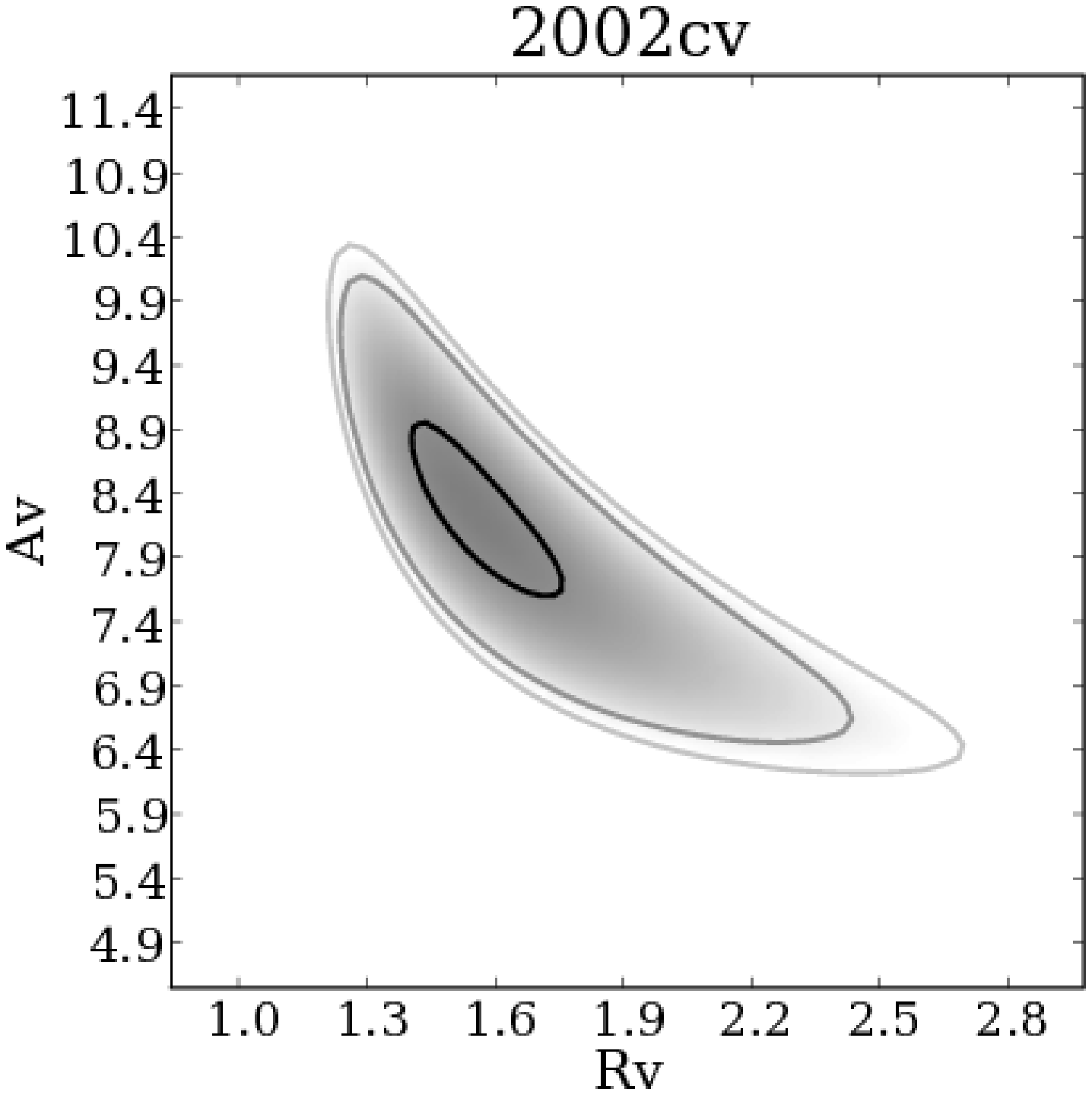,width=9cm,height=5.5cm}
\psfig{figure=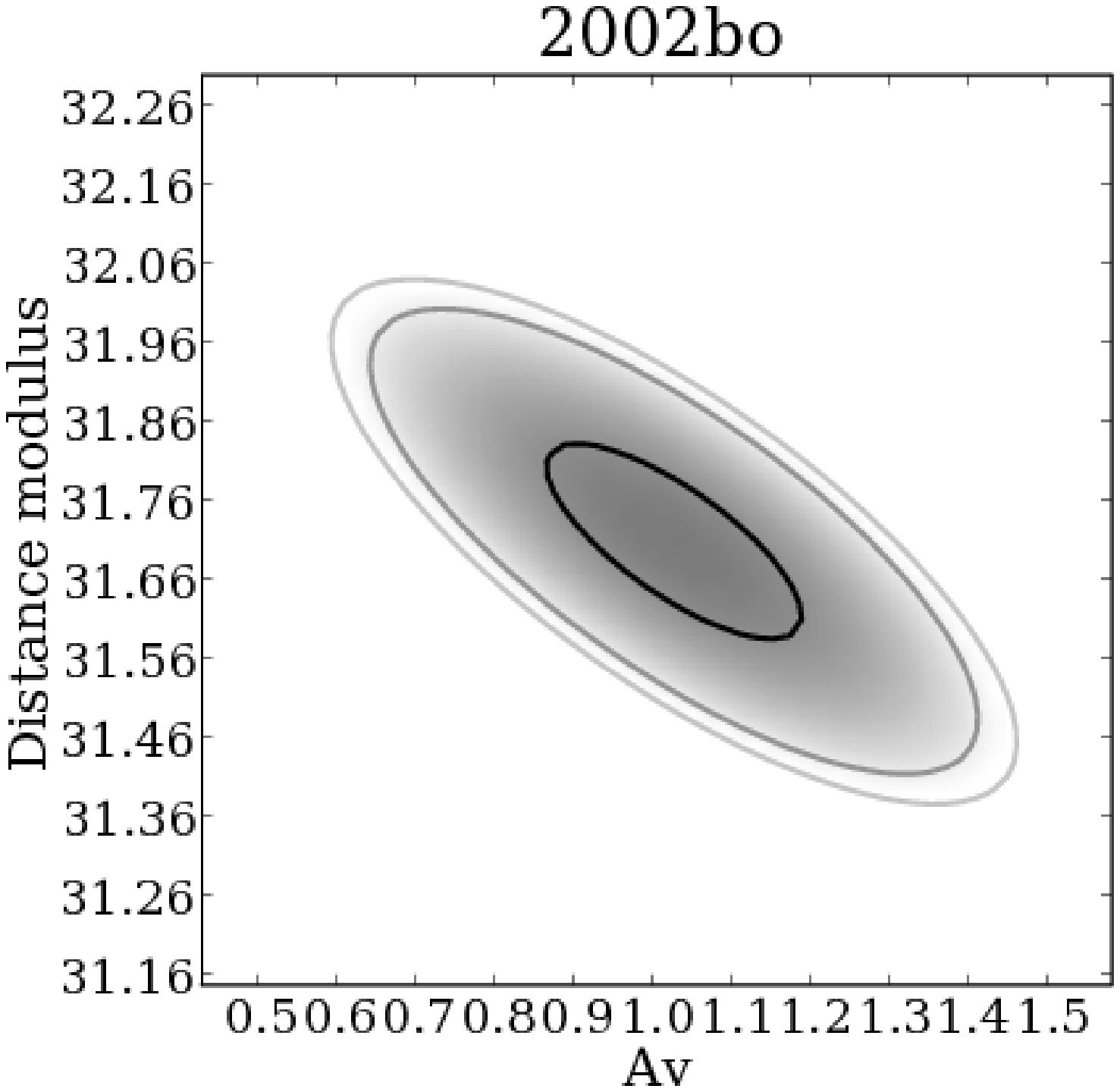,width=9cm,height=5.5cm}
\caption{Multi-dimensional maximum-likelihood estimates to derive
\av\ and \rv\ toward SN~2002cv and SN~2002bo along with the
host-galaxy distance. Each contour corresponds to 3$\sigma$ (outer),
2$\sigma$ (middle), and 1$\sigma$ (inner). The values found were
\av\ = 8.40 $\pm$ 0.35 mag and \rv\ = 1.60 $\pm$ 0.10 for SN~2002cv,
\av\ = 1.00 $\pm$ 0.10 mag for SN~2002bo, and $\mu$ = 31.76 $\pm$
0.07 mag.}
\label{fig_contours}
\end{center}
\end{figure}

We note that the value of $\mu$ found here is in excellent
agreement with $\mu$ = 31.77 mag, the distance modulus derived using
the relative distance of NGC 3190 from the Virgo cluster (1.48 mag,
\citealt{kraan86}), and assuming a Virgo cluster distance of 15.3~Mpc
\citep{freedman01}. An alternative estimate of the distance is
obtained considering that NGC~3190 is a member of the Leo~III
group \citep{garcia93}; the surface brightness fluctuation
\citep{tonry01a} distance of another possible member of the
group, NGC~3226, is $\mu$ = 31.86 $\pm$ 0.24 mag \citep{krisc04b}.
This is also in good agreement with our estimate above, which we
hereafter adopt.\\

In summary, the \av\ and \rv\ estimates obtained with the three
different procedures are listed in Table \ref{tabla_reddvalue}.
The weighted averages of these values are \av\ = 8.66 $\pm$ 0.21
mag and \rv\ = 1.59 $\pm$ 0.07.

Including the Galactic extinction component, \ebv$_{Gal}$ = 0.025 mag
\citep{schlegel98}, the total extinction suffered by SN~2002cv is
\av$_{tot}$ = 8.74 $\pm$ 0.21 mag, making SN~2002cv one of the most
highly extinguished SNe~Ia ever observed.\\

\begin{table}
\centering \caption{Values of \av\ and \rv\ derived from
different methods.}\label{tabla_reddvalue}
\begin{tabular}{ccc}
\hline
 Method           &   $A_{V,host}$ (mag)  &  \rv\ \\
\hline
Colour Evolution  & $8.99 \pm 0.30$ & $1.97 \pm 0.30$ \\
Comp-CCM          & $8.17 \pm 0.57$ & $1.52 \pm 0.11$ \\
Multi-dimensional & $8.40 \pm 0.35$ & $1.60 \pm 0.10$ \\
\hline
\end{tabular}
\end{table}

With regard to the extinction law, we stress that all three
methods give a low value of \rv. The small value of \rv\
obtained for this source, which is deeply embedded in a dust
lane, appears consistent with the scenario proposed by
\cite{goudfrooij94} and \cite{patil06}, who suggest that the
observed dust-grain size in the host-galaxy dust lanes may be altered
by different mechanisms, such as destruction of grains due to
sputtering in supernova blast waves, grain-grain collisions, or
sputtering by warm and hot thermal ions \citep{goudfrooij94}. A
possible different mechanism for producing small dust grains is
erosion by the SN radiation field \citep{whittet92}. In this case,
the dust must be located close to the SN, possibly even originating
in the progenitor evolution.

We should mention, however the alternative explanation for the
small apparent value of \rv\ proposed  by \cite{wangl05}, which
invokes the effect of a light echo from circumstellar dust. In
this scenario, the physical properties of the dust are much less
important than its distribution in the immediate neighborhood of
SNe~Ia.

In conclusion, SNe~Ia can be a very useful tool for studying the
properties of dust in distant galaxies.

\section{Photometric parameters} \label{param}

The parameters characterizing the photometric behaviour of SNe~Ia
are usually derived from $B$ and $V$ light curves. For SN~2002cv,
the $B$ and $V$ bands are heavily extinguished and no measurements
were obtained (we measured only upper limits in the $V$ band).
Hence, to characterize this SN with respect to other events we
will seek general correlations between blue to red light-curve
parameters of SNe~Ia.

\subsection{$VRI$ decline rates vs. \dm15}
\label{param_rel_vri}

\cite{hamuy96c}, using a sample of seven SNe~Ia, showed a
strong correlation between \dm15 and other light-curve
parameters, namely $\Delta m_{60}$(B), $\Delta m_{20}$(V), $\Delta
m_{60}$(V), and $\Delta m_{60}$(I). They also found that for the $I$
band, the time interval between the minimum and the secondary
maximum is greater for the slowly declining SNe~Ia.

\begin{figure}
\begin{center}
\psfig{figure=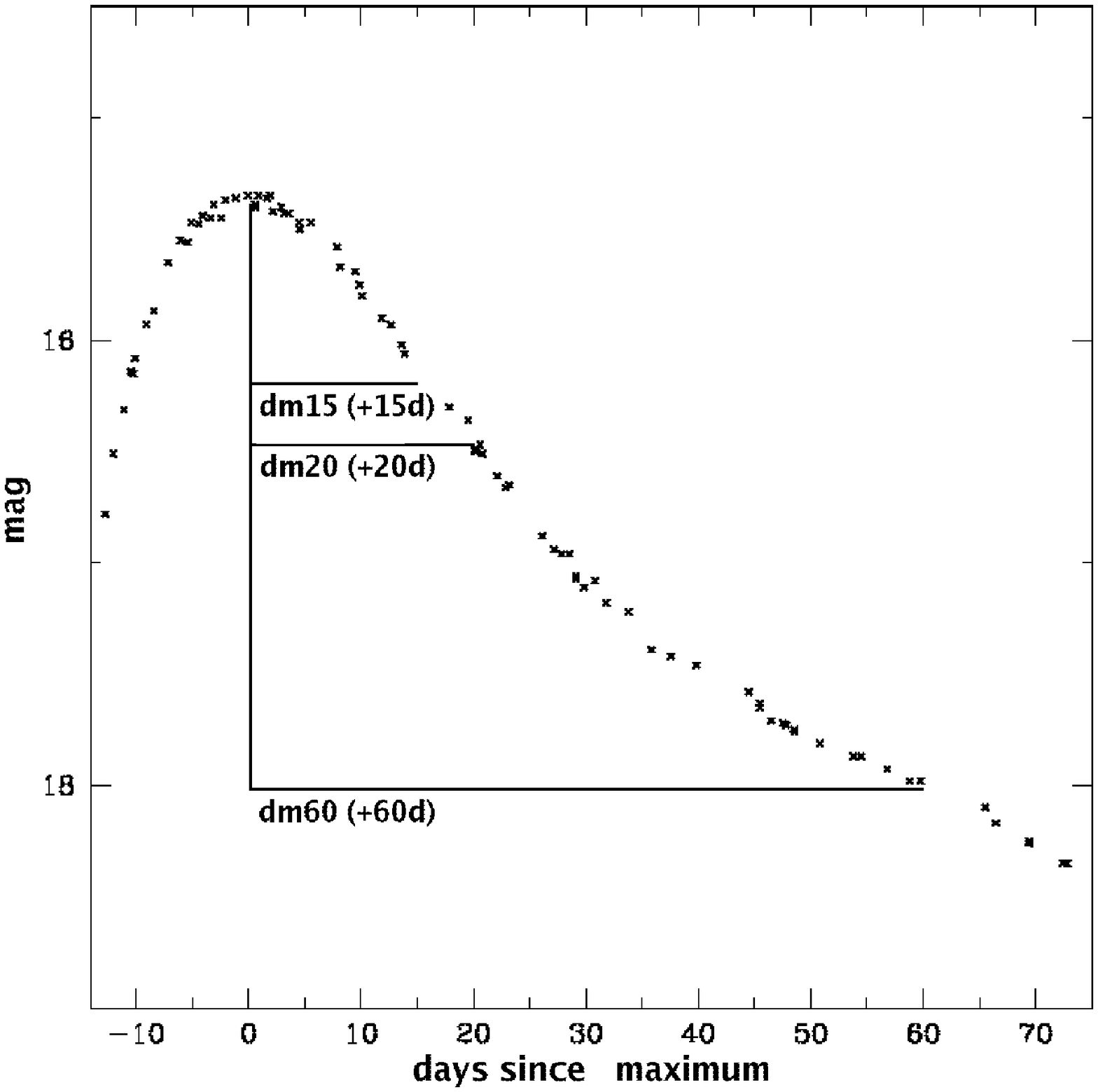,width=8cm,height=8cm}
\psfig{figure=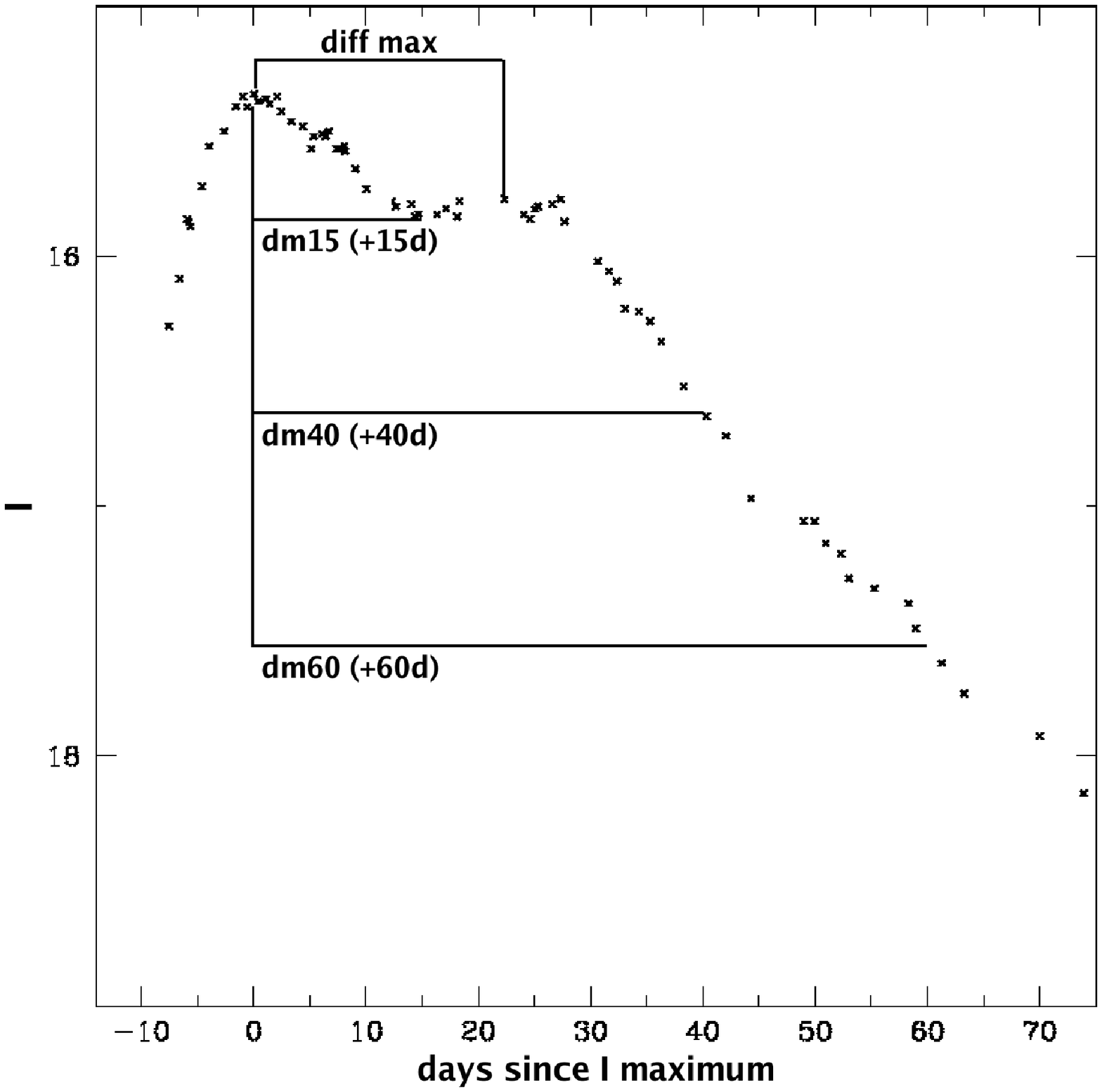,width=8cm,height=8cm}
\caption{Graphical representation of the parameters defined in
Section \ref{param} for generic bands (top panel) and in
particular for the $I$ band (bottom panel).} \label{fig_graprepr}
\end{center}
\end{figure}

Here, we extend the statistics of \cite{hamuy96c} to a sample of
20 SNe~Ia (Table \ref{tabla_mainsne}) spanning a range in \dm15
between 0.90 and 1.94 mag. For each SN we obtained the epoch and
magnitude of the maximum light in the $V$, $R$, and $I$ bands, and
measured the decay in the first 15, 20, 40, and 60 days after
maximum in each band. See Figure \ref{fig_graprepr} for a graphical
description of these parameters.

We note that the decline rates presented here are not corrected
for extinction. \cite{phillips99} showed that \dm15 has a weak
dependence on reddening. To check if a similar correction was
required in the other bands, we measured \dm15, the decline rates
$\Delta m$ in other bands, as well as \dijd\ using the spectrophotometry
of three SNe~Ia (SN~1994D, SN~1996X, and SN~2004eo). The measurements
were performed for (a) the unreddened spectra and (b) the spectra
reddened by the Galactic and host-galaxy extinction with the laws
found for SN~2002cv (see Section \ref{redd}). We confirm the
correction for \dm15 given by \cite{phillips99}, but the
corrections for the $I$-band parameters turned out to be negligible.

\begin{figure}
\begin{center}
\psfig{figure=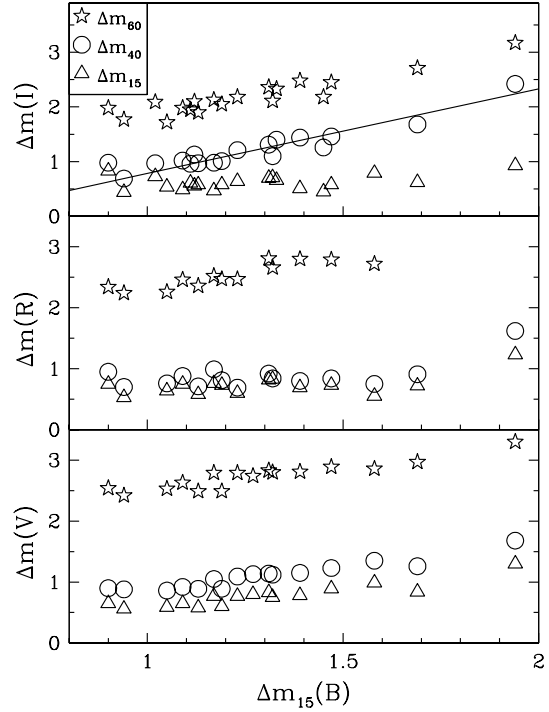,width=8.5cm,height=11.3cm}
\caption{$VRI$ light-curve decline rates $\Delta m_{15}$
(triangles), $\Delta m_{20}$ (circles), and $\Delta m_{60}$ (stars)
vs. $\Delta m_{15}(B)_{obs}$. The straight line is a fit to the
$\Delta m_{40}(I)$ points.} \label{fig_dm}
\end{center}
\end{figure}

Figure \ref{fig_dm} (bottom) shows the decline rates of the $V$, $R$,
and $I$ light curves versus \dm15. The correlations between the
different decline rates with \dm15 are similar to those found by
\cite{hamuy96c}.

The behaviour of the $R$-band decline rates (Figure \ref{fig_dm},
middle) is different according to the different intervals
considered. While there are no correlations between \dm15 and
$\Delta m_{15}(R)$ or $\Delta m_{20}(R)$ except for rapidly declining
SNe (SN~1991bg), we find a stronger correlation with $\Delta
m_{60}(R)$. This behaviour of the decline rates is probably due to
the change of opacity and concentration of iron-peak elements in
the central regions during these first days \citep{kasen06}, which
produce the secondary maximum. According to this model, a SN~Ia
with homogeneous abundance distribution shows red light curves in
which the first and second maxima are indistinguishable. This
could be the case for SN~1991bg, which has no secondary maximum
and exhibits fast decline in the $B$ band and in redder bands.

In Figure \ref{fig_dm} (top) we also compare the parameters
measured for the $I$-band light curve with \dm15. While it appears that
$\Delta m_{15}(I)$ remains constant, there is a clear correlation
between $\Delta m_{40}(I)$ and $\Delta m_{60}(I)$ and \dm15.

In the $I$ band, the secondary maxima are very pronounced, as well
as in the $J$, $H$, and $K$ light curves. The strength and phase of the
secondary maximum is found to correlate with \dm15, being more
prominent and delayed in luminous SNe~Ia \citep{hamuy96c,nobili05}.
We measured the magnitude difference $\Delta I_{max}$ and time
interval \dijd\ between the primary and secondary peaks in the
$I$-band light curve for the SNe of our sample (see Figure
\ref{fig_graprepr}), and compared them with \dm15. Note that SNe
with \dm15 $\geq$ 1.8 mag, similar to SN~1991bg, are not included in
the diagram because their $I$ light curves do not show a secondary
maximum.

\begin{figure}
\begin{center}
\psfig{figure=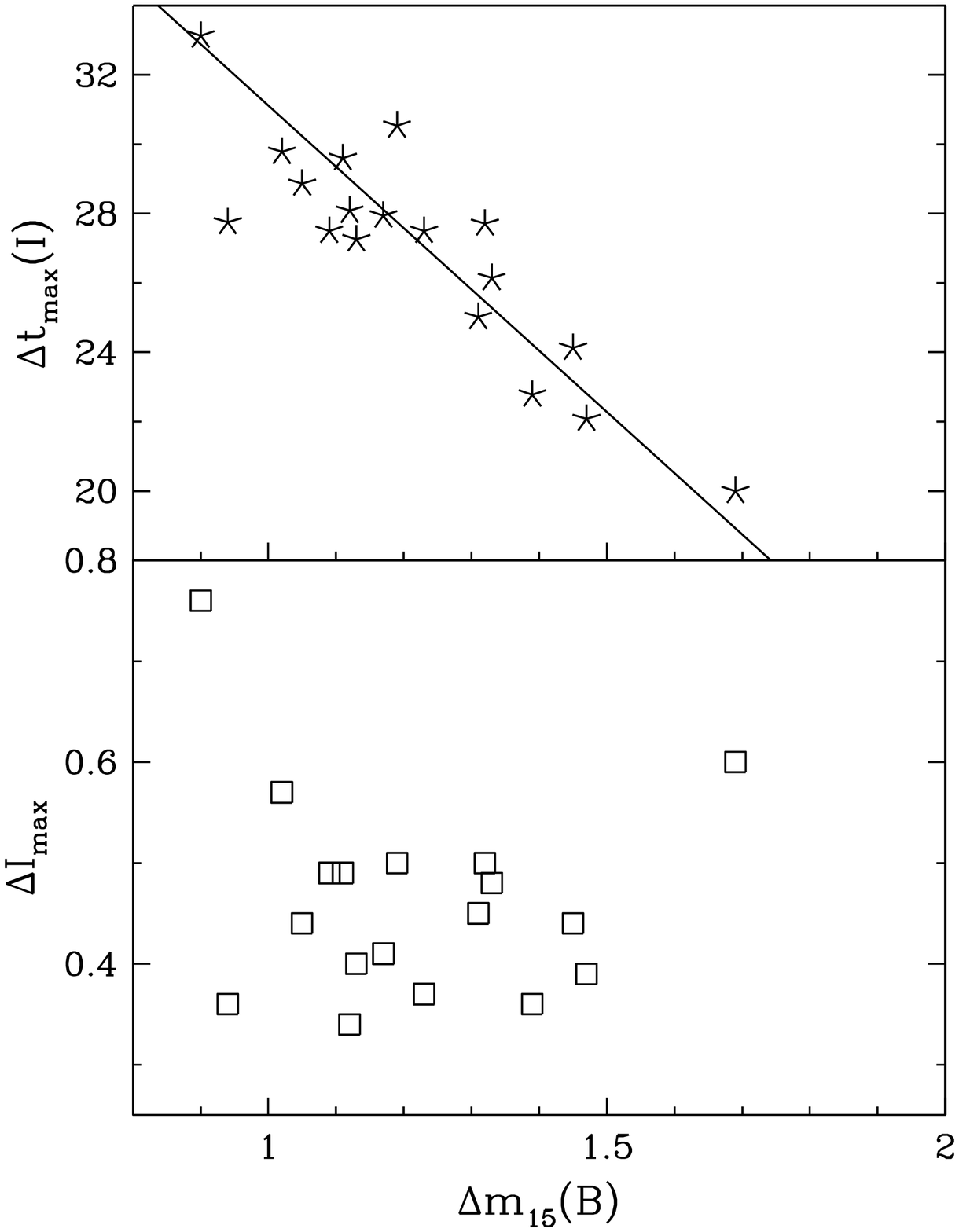,width=8.5cm,height=11.3cm}
\caption{Difference in phase (top) and of magnitude (bottom)
between the primary and secondary maxima of the $I$-band light
curve vs. $\Delta m_{15}(B)_{obs}$. The straight line is the
best fit to the \dijd\ points.} \label{fig_diffImag}
\end{center}
\end{figure}

While $\Delta I_{max}$ shows no clear correlation with \dm15
(Figure \ref{fig_diffImag}, bottom), the correlation between
\dijd\ and \dm15 is tight (Figure \ref{fig_diffImag}, top), with
the phase delay of the secondary maximum being longer for
slowly declining SNe~Ia. \\

\subsection{$I$-band magnitudes and \dm15} \label{param_rel_idm15}

\begin{figure}
\begin{center}
\psfig{figure=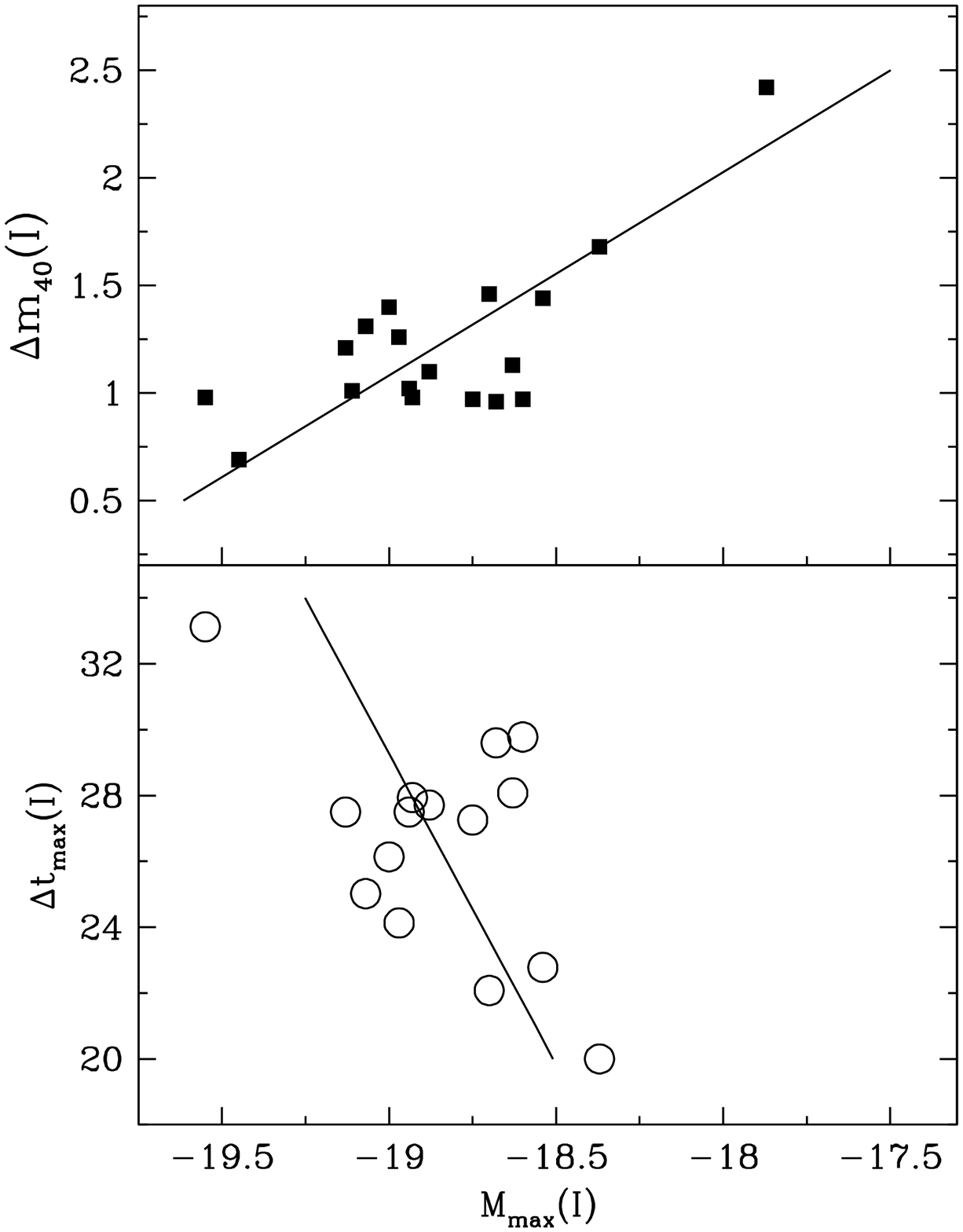,width=8.5cm,height=11.3cm}
\caption{$\Delta m_{40}(I)$ (top) and \dijd\ (bottom) vs.
$M_I^{max}$, for the SNe~Ia of Table \ref{tabla_mainsne}. The
straight lines are the best fits.} \label{fig_absmag}
\end{center}
\end{figure}

It is well known that the early-time decline rate in the $B$-band
light curve of SNe~Ia correlates with the luminosity. Since the $B$
band is not available for SN~2002cv, we are forced to find alternative
indicators to establish the photometric subclass of this SN. In
Figure \ref{fig_lcdm15} we have already shown that the $I$ light-curve
shapes of SNe 2002cv and 1992A were similar, so it is likely that
they have similar \dm15 $\approx 1.5$ mag.

In order to obtain a more
accurate estimate of \dm15, we exploit the previously derived
correlations with $\Delta m_{40}(I)$ (SN~2002cv was observed until
+44.2 days after $I$ maximum) and \dijd. Using the code developed by
\cite{akritas96} for linear regression analysis, we performed a
linear fit to the points in Figures \ref{fig_dm} (top) and
\ref{fig_diffImag} (top), obtaining (respectively)

\begin{equation}
  \Delta t_{max}(I) =  (48.80 \pm 1.89) - (17.68 \pm 1.52) \times
\Delta m_{15}(B),
\label{equ_dijd_dm15}
\end{equation}

\noindent
and

\begin{equation}
  \Delta m_{40}(I) = (1.55 \pm 0.17) \times \Delta m_{15}(B) -
(0.77 \pm 0.23).
\label{equ_dmi40_dm15}
\end{equation}

\noindent
For equation \ref{equ_dijd_dm15}, we used 16 SNe~Ia and the Pearson
correlation coefficient is $-0.94$; for equation
\ref{equ_dmi40_dm15}, we used 18 SNe~Ia and the Pearson correlation
coefficient is 0.93\footnote{In the fit of \dijd\ vs. \dm15 we
excluded SN~1991T which appears to deviate (as in
\citealt{hamuy96c}). Indeed, SN~1991T was an abnormal object with a
number of pre-maximum spectroscopic peculiarities
\citep{filippenko92a,phillips92}}.\\

Since \dm15 is related to the $B$ luminosity of SNe~Ia, we
could expect a similar relation between $\Delta m_{40}(I)$ and
$M_I^{max}$, the absolute magnitude of the first $I$ maximum, and
also between \dijd\ and $M_I^{max}$. Figure \ref{fig_absmag}
confirms this expectation, though in both cases the dispersion is
high mainly due to the uncertainties in the distance moduli used
for computing $M_I^{max}$.

The linear fits give

\begin{equation}
  M_I^{max} = (-0.05 \pm 0.02) \times \Delta t_{max}(I) - (17.45 \pm 0.68),
\label{equ_absmag_dijd}
\end{equation}

\noindent
and

\begin{equation}
  M_I^{max} = (1.06 \pm 0.24) \times \Delta m_{40}(I) - (20.14 \pm 0.31).
\label{equ_absmag_dmi40}
\end{equation}

\noindent
For the two relations 15 and 18 SNe~Ia were used, and the
correlation coefficients are 0.57 and 0.73, respectively.\\


\begin{table*}
\centering \caption{Main parameters for the SN~Ia sample.}
\label{tabla_mainsne}
\begin{tabular}{lllllllll}
\hline
SN & $\Delta m_{15}(B)_{obs}$ & JD($I_{max}$) $-$ & $I_{max}$ & $\Delta m_{40}(I)_{obs}$ & $\Delta max(I)$ & $\mu$ & $A_{I,tot}^{\ast}$ & sources\\
   &                          & 2,400,000.00    &         &   &   &   &   &   \\
\hline
SN~1992bc & 0.90(0.05) & 48910.84(0.50) & 15.56(0.05) & 0.98(0.06) & 33.13(1.13) & 34.56 & 0.04 & 0,1,2,3\\
SN~1991T  & 0.94(0.05) & 48371.60(0.50) & 11.62(0.04) & 0.69(0.01) & 27.75(0.51) & 30.74 & 0.33 & 0,1,2,4,5,6,7\\
SN~2003du & 1.02(0.05) & 52764.92(0.50) & 13.83(0.02) & 0.97(0.03) & 29.78(0.71) & 32.42 & 0.01 & 0,1,8\\
SN~1990N  & 1.05(0.05) & 48080.70(0.50) & 12.95(0.02) & -- & 28.86(1.05) & 31.07 & 0.10 & 0,1,2,6\\
SN~1992al & 1.09(0.05) & 48836.86(0.50) & 14.93(0.04) & 1.02(0.02) & 27.50(0.71) & 33.82 & 0.05 & 0,1,2,3\\
SN~2005cf & 1.11(0.03) & 53532.00(0.50) & 13.70(0.03) & 0.96(0.03) & 29.60(0.55) & 32.19 & 0.14 & 0,1,9\\
SN~2003cg & 1.12(0.05) & 52728.22(0.50) & 13.82(0.04) & 1.13(0.04) & 28.08(0.71) & 31.61 & 1.14 & 0,1,10\\
SN~2001el & 1.13(0.05) & 52181.43(1.00) & 12.81(0.04) & 0.97(0.04) & 27.26(1.01) & 31.29 & 0.27$\dagger$ & 0,1,2,11 \\
SN~2002bo & 1.17(0.05) & 52355.50(1.00) & 13.49(0.10) & 0.98(0.10) & 27.94(1.00) & 31.77 & 0.56 & 0,1,12,13\\
SN~1995E  & 1.19(0.05) & 49771.00(0.50) & 15.33(0.05) & 1.01(0.05) & -- & 33.43 & 1.01 & 0,1,2,14\\
SN~1998dh & 1.23(0.17) & 51029.32(0.50) & 14.10(0.05) & 1.21(0.05) & 27.50(0.69) & 32.92 & 0.31 & 0,1,2,15\\
SN~1994D  & 1.31(0.05) & 49428.50(1.00) & 12.11(0.05) & 1.31(0.06) & 25.02(1.01) & 31.14 & 0.04 & 0,1,16,17,18,19\\
SN~1996X  & 1.32(0.05) & 50188.00(1.00) & 13.39(0.01) & 1.10(0.02) & 27.71(1.05) & 32.17 & 0.10 & 0,1,14,20\\
SN~2002er & 1.33(0.04) & 52523.56(0.50) & 14.49(0.05) & 1.40(0.09) & 26.14(0.71) & 32.90 & 0.53 & 0,1,21\\
SN~1997E  & 1.39(0.06) & 50466.19(0.50) & 15.46(0.05) & 1.44(0.07) & 22.78(1.41) & 33.72 & 0.28 & 0,1,2,15\\
SN~2004eo & 1.45(0.04) & 53276.30(1.00) & 15.36(0.04) & 1.26(0.04) & 24.13(1.03) & 34.12 & 0.21 & 0,1,22\\
SN~1992A  & 1.47(0.05) & 48638.00(0.50) & 12.80(0.04) & 1.46(0.05) & 22.08(0.56) & 31.41$\ddagger$ & 0.09 & 0,1,2,7,23\\
SN~2000cn & 1.58(0.12) & 51706.93(0.50) & 16.63(0.05) & -- & -- & 34.93 & 0.16 & 0,1,2,15\\
SN~1992bo & 1.69(0.05) & 48984.94(0.50) & 15.95(0.05) & 1.68(0.05) & 20.00(0.71) & 34.28 & 0.08 & 0,1,2,3\\
SN~1991bg & 1.94(0.05) & 48608.60(1.00) & 13.51(0.05) & 2.42(0.05) & -- & 31.32 & 0.06 & 0,1,2,24,25,26,27\\
\hline
\end{tabular}
\begin{flushleft}
0 = This work; 1 = LEDA; 2 = \cite{reindl05}; 3 = \cite{hamuy96b};
4 = \cite{schmidt94}; 5 = \cite{cappellaro97}; 6 = \cite{lira98};
7 = \cite{altavilla04}; 8 = \cite{stanishev06}; 9 =
\cite{pastorello06a}; 10 = \cite{elias06}; 11 = \cite{krisc03}; 12
= \cite{benetti04}; 13 = \cite{krisc04a}; 14 = \cite{riess99}; 15
= \cite{jha06a}; 16 = \cite{richmond95}; 17 = \cite{tsvetkov95};
18 = \cite{patat96}; 19 = \cite{meikle96}; 20 = \cite{salvo01}; 21
= \cite{pignata04a}; 22 = \cite{pastorello06b}; 23 =
\cite{suntzeff96}; 24 = \cite{filippenko92b}; 25 =
\cite{leibundgut93}; 26 = \cite{turatto96}; 27 = \cite{tonry01b};
$^{\ast}$ $A_{I,Gal}$ + $A_{I,host}$; $^{\dagger}$ $E(B-V)_{host}$ =
0.18 mag, with $R_V$ = 2.88; $^{\ddagger}$ average $\mu$ from
different sources.\\
\end{flushleft}
\end{table*}

\begin{table}
\centering \setlength\tabcolsep{3pt} \caption{Main data of
SN~2002cv and its host galaxy.}\label{tabla_mainphdata}
\setlength\tabcolsep{2pt} 
\begin{tabular}{lll}
\hline
 Host-Galaxy Data & NGC 3190 &  Ref.\\
\hline
$\alpha$ (2000)   & $10^{h}18^m 05\psec60$ & 1 \\
$\delta$ (2000)   & $+21^{\circ} 49\arcmin 55\arcsec$& 1\\
Galaxy type       & SA(s)a pec & 1\\
B magnitude       & 12.12& 1 \\
\ebv$_{Gal}$      & 0.025 & 2\\
$v_{r, helio}$ (\kms)$^*$& 1271 & 1\\
$\mu$             & 31.76 $\pm$ 0.07 & 3 \\
\hline \hline
 SN Data          & SN~2002cv  &  Ref.\\
\hline
$\alpha$ (2000)   & $10^{h} 18^{m} 03\psec68$ & 4 \\
$\delta$ (2000)   & $+21^{o} 50\arcmin 06\parcsec20$& 4 \\
Offset SN-Gal. nucleus & $18\arcsec W, 10\arcsec N$ & 4 \\
Discovery date (UT)  & 2002 May 13.7 & 4 \\
Discovery date (JD)  & 2452408.20 & 4 \\
\ebv$_{host}$     & 5.45 $\pm$ 0.28 & 3 \\
\rv$_{host}$      & 1.59 $\pm$ 0.07 & 3 \\
\av$_{tot}$       & 8.74 $\pm$ 0.21 & 3 \\
Date of $I$ max (JD)& 2,452,415.09 $\pm$ 0.22 & 3 \\
Magnitude and epoch & $V < 19.70$;  $\sim$+5.8 (days) & 3 \\
at max wrt $I$ max & $R$ = 19.08 $\pm$ 0.20;   +3.2 (days) & 3 \\
                   & $I$ = 16.57 $\pm$ 0.10;   0.0 (days) & 3 \\
                   & $J$ = 14.75 $\pm$ 0.03;   -0.4 (days) & 3 \\
                   & $H$ = 14.34 $\pm$ 0.01;  -2.0 (days) & 3 \\
                   & $K$ = 13.91 $\pm$ 0.04;  +1.6 (days) & 3 \\
Magnitude and epoch & $I$ = 17.11 $\pm$ 0.10;  +23.1 (days) & 3 \\
of secondary $IJHK$  & $J$ = 15.24 $\pm$ 0.19;  +41.5 (days) & 3 \\
max wrt $I$ max      & $H$ = 14.40 $\pm$ 0.11;  +25.9 (days) & 3 \\
                    & $K$ = 13.88 $\pm$ 0.29;  +35.2 (days) & 3 \\
Estimated $\Delta m_{15}(B)_{intr}$ & 1.46 $\pm$ 0.17  & 3 \\
Absolute magnitude & $M_R^{max} = -19.05 \pm 0.27$ & 3 \\
(derived from the distance)& $M_I^{max} = -18.79 \pm 0.20$ & 3 \\
                   & $M_J^{max} = -18.50 \pm 0.12$ & 3 \\
                   & $M_H^{max} = -18.41 \pm 0.09$ & 3 \\
                   & $M_K^{max} = -18.44 \pm 0.09$ & 3 \\
log$_{10}L$ ($RIJHK$)& 42.54 $\pm$ 0.30 \ergs & 3 \\
log$_{10}L_{{\it uvoir}}$ & 42.94 $\pm$ 0.60 \ergs & 3 \\
M($^{56}$Ni)       & 0.44$^{+0.65}_{-0.26}$ (\M) & 3 \\
\hline
\end{tabular}
\begin{flushleft}
$^*$Heliocentric radial velocity.\\
(1) NED; (2) \cite{schlegel98}; (3) this work; (4) \cite{lavionov02}. \\
\end{flushleft}
\end{table}

\subsection{Main photometric parameters of SN~2002cv} \label{param_value}

The maximum-light epochs and magnitudes for the different bands,
reported in Table \ref{tabla_mainphdata}, were derived by fitting
low-order polynomials (lower than 5 deg) to the light curves. The
first $I$ maximum is found on JD($I_{max}$) = 2,452,415.09 $\pm$ 0.22
(May 20.6, 2002 UT) with $I_{max} = 16.57 \pm 0.10$ mag. The $I$-band
secondary maximum occurred \dijd\ = 23.06 $\pm$ 0.59 days later,
with magnitude 17.11 $\pm$ 0.10. We also measured $\Delta
m_{40}(I) = 1.50 \pm 0.32$ mag.

Using equations \ref{equ_dijd_dm15} and \ref{equ_dmi40_dm15}, we
derived an average value \dm15\ = 1.46 $\pm$ 0.17, indeed very
close to that of SN~1992A and SN~2004eo.

Hereafter, from the measured \dijd\ and $\Delta m_{40}(I)$, and
equations \ref{equ_absmag_dijd} and \ref{equ_absmag_dmi40}, we
derive an estimate of the intrinsic $I$ absolute magnitude at
maximum for SN~2002cv. The values obtained are $M_I^{max} =
-18.60 \pm 0.84$ mag and $M_I^{max} = -18.55 \pm 0.76$ (respectively),
with a weighted average value of $<M_I^{max}> = -18.57 \pm
0.57$ mag. This magnitude is close to the average value for SNe~Ia
found by \cite{saha99}: $<~M_I^{max}> = -18.74 \pm 0.03$ mag (scaled to
$H_0$ = 72 \kms $Mpc^{-1}$).

For the $I$ band we can also derive the absolute magnitude through
the relation with \dm15. Using (a) the known relation between the
peak luminosity and the \dm15 of \cite{hamuy96a} (``peak subsample"
case), (b) \cite{phillips99} (see their Table 3), and (c)
\cite{prieto06} (complete sample reported in Table 3), scaled to
$H_0$ = 72 \kms $Mpc^{-1}$, we derived $M_I^{max}$ quite close to
that estimated with our equations (see Table \ref{tabla_absolmag}).
We note that equation \ref{equ_magabs} with the parameters
previously derived provide $M_I^{max} = -18.79 \pm 0.20$ mag,
in excellent agreement with the above estimate within the
calculated errors. Note that we consider the absolute magnitudes
of SN~2002cv to be those obtained using the distance modulus.

Considering again the \cite{prieto06} relation, we obtain an
analogous estimate in the $R$ band, $M_R^{max} = -19.04 \pm 0.13$ mag.
\cite{hamuy96a} and \cite{phillips99} do not provide an average
relation in the $R$ band. As before, by applying equation
\ref{equ_magabs}, we obtain a value in excellent agreement,
$M_R^{max} = -19.05 \pm 0.27$ mag.

The NIR absolute magnitudes were estimated from the observed peak
magnitude and $\mu$. Their values ($M_J^{max} = -18.50
\pm 0.12$ mag, $M_H^{max} = -18.41 \pm 0.09$ mag, and $M_K^{max} =
-18.44 \pm 0.09$ mag) are consistent with the average values found by
\cite{krisc04b} within the uncertainties ($M_J^{max} = -18.61 \pm
0.13$ mag, $M_H^{max} = -18.28 \pm 0.15$ mag, and $M_K^{max} = -18.44
\pm 0.14$ mag).

\begin{table}
\centering \setlength\tabcolsep{3pt} \caption{Absolute $I$ magnitude
of SN~2002cv ($H_0$ = 72 \kms $Mpc^{-1}$).}\label{tabla_absolmag}
\setlength\tabcolsep{2pt}
\begin{tabular}{cll} \hline
$M_I^{max}$ (mag)  & Method &  Reference \\
\hline
$-18.79 \pm 0.20$ & distance & this work\\
\hline
$-18.57 \pm 0.57$ & eq. \ref{equ_absmag_dijd} and \ref{equ_absmag_dmi40} & this work \\
\hline
$-18.52 \pm 0.20$ & $M_I^{max}$ vs. \dm15 & \cite{hamuy96a}$^1$    \\
$-18.52 \pm 0.28$ & $M_I^{max}$ vs. \dm15 & \cite{phillips99}$^2$   \\
$-18.79 \pm 0.12$ & $M_I^{max}$ vs. \dm15 & \cite{prieto06}$^3$     \\
\hline
\end{tabular}
\begin{flushleft}
($1$) According to the relation given in Table 3 of
\cite{hamuy96a} (peak luminosity);\\
($2$) according to the relation given in Table 3 of
\cite{phillips99}; \\
($3$) according to the relation given in Table 3 of
\cite{prieto06} (for the complete sample -- ALL).
\end{flushleft}
\end{table}

\section{Summary} \label{summary}

We have presented optical and NIR data on SN~2002cv and
completed the study of this SN started by \cite{dipaola02}
including the available data collected worldwide. SN~2002cv
exploded in the same galaxy as SN~2002bo, a few months later, and
it suffered a very high extinction.
The presence of clear secondary maxima in the red and NIR light
curves, the colour curves, and the spectral features confirm its
early classification as a SN~Ia.

We have used different methods to estimate the extinction, adopting in
all cases a CCM extinction law with \rv\ as a free parameter. We
obtained \ebv\ = 5.45 $\pm$ 0.28 mag and \rv\ = 1.59 $\pm$ 0.07 ($A_{V
host}$ = 8.66 $\pm$ 0.21) inside the host galaxy, which give a
total absorption \av\ = 8.74 $\pm$ 0.21 mag. It turns out that
SN~2002cv is one of the most reddened SNe ever observed, and
a new entry in the growing list of SNe with a low value of \rv\
(e.g., SN~1999cl, \citealt{krisc06a}; SN~2001el,
\citealt{krisc07}; SN~2002hh, \citealt{pozzo06}; SN~2003cg,
\citealt{elias06}).

We have constructed empirical relations between \dm15 and
$M_I^{max}$, the delay of the secondary $I$ maximum (\dijd), and the
decline rate at 40 days in the $I$ band ($\Delta m_{40}(I)$). These
allow us to derive the luminosity class of a SN~Ia, without the
use of the unavailable $B$-band data. With these average relations
for SN~2002cv, we derive $\Delta m_{15}(B)_{true} = 1.46 \pm
0.17$ mag and $M_I^{max} = -18.57 \pm 0.57$ mag, which is marginally
fainter than that estimated from the observed peak magnitude and
$\mu$: $M_I^{max} = -18.79 \pm 0.20$ mag.

Since \rv\ is related to the grain size, the different values of
this found for SN~2002bo and SN~2002cv show that in NGC~3190 there
is dust with different properties (at least for the two line of
sights measured) which are different from those of typical
Galactic dust. A possible explanation for the different grain
sizes might be the effect of SN radiation on local circumstellar
dust or the peculiar physical conditions in dust lanes.

A better understanding of this phenomenon is important both for
understanding the nature of the exploding stars and for the
calibration of SNe~Ia for cosmological use.\\

\bigskip

\noindent {\bf ACKNOWLEDGMENTS}

This work is based on observations collected at the European
Southern Observatory, Chile, under programmes ID 69.D-0672(B) and
ID 069.D-0672(C), at the Isaac Newton and Jacob Kapteyn 1.0~m
Telescopes of the Isaac Newton Group (La Palma); the Asiago 1.82~m
Copernico Telescope; the Teramo-Normale Telescope and the AZT-24
Telescope (INAF Observatories, Italy); the 0.76~m Katzman
Automatic Imaging Telescope, the Nickel 1~m telescope, and the
Shane 3~m reflector (Lick Observatory, California, USA); and the
United Kingdom Infrared Telescope (Hawaii). The INT and JKT are
operated on the island of La Palma by the Isaac Newton Group (ING)
in the Spanish Observatorio del Roque de los Muchachos of the
Instituto de Astrof\'isica de Canarias. The European observations
were obtained in the framework of the European Supernova
Collaboration
(ESC)\footnote{http://www.mpa-garching.mpg.de/$\sim$rtn/} funded
as a European Research Training Network (HPRN-CT-2002-00303).
AZT-24 Telescope is installed in Campo Imperatore (Italy) under
agreement between Pulkovo, Rome and Teramo observatories. KAIT was
constructed and supported by donations from Sun Microsystems,
Inc., the Hewlett-Packard Company, AutoScope Corporation, Lick
Observatory, the US National Science Foundation (NSF), the
University of California, the Sylvia \& Jim Katzman Foundation,
and the TABASGO Foundation. UKIRT is operated by the Joint
Astronomy Centre on behalf of the U.K. Particle Physics and
Astronomy Research Council. Some of the data reported here were
obtained as part of the ING and UKIRT Service Programmes. A.V.F.'s
supernova group at U.C. Berkeley is funded by NSF grant
AST--0607485 and the TABASGO Foundation. This work has made use of
the NASA/IPAC Extragalactic Database (NED), which is operated by
the Jet Propulsion Laboratory, California Institute of Technology,
under contract with the National Aeronautics and Space
Administration.

\end{document}